\documentclass{aa}  

\usepackage[utf8]{inputenc}
\usepackage{amssymb,amsmath}
\usepackage{graphicx}
\usepackage[dvipsnames]{xcolor}
\usepackage[breaklinks=true]{hyperref}
\hypersetup{colorlinks=true,allcolors=teal}
\usepackage[flushleft]{threeparttable}
\usepackage{footmisc}
\usepackage{subfigure}
\usepackage{multirow}

\newcommand{\be}{\begin{equation}}
\newcommand{\ee}{\end{equation}}
\def\bear#1\ear{\begin{align}#1\end{align}}

\usepackage[T1]{fontenc}
\usepackage{amsmath}	
\DeclareRobustCommand{\VAN}[3]{#2}
\let\VANthebibliography\thebibliography
\def\thebibliography{\DeclareRobustCommand{\VAN}[3]{##3}\VANthebibliography}
\usepackage{graphicx}
\usepackage{txfonts}
\begin{document}

   \title{Efficient Modelling of Lyman-$\alpha$ opacity fluctuations during late reionization epoch}
    \authorrunning{Maity et al.}

   \author{Barun Maity\inst{1}\fnmsep\thanks{maity@mpia.de}
          ,
          Frederick Davies
          \inst{1},
          \and
          Prakash Gaikwad
          \inst{2}}

   \institute{Max-Planck-Institut f\"ur Astronomie, K\"onigstuhl 17, D-69117 Heidelberg, Germany
         \and Department of Astronomy, Astrophysics and Space Engineering, Indian Institute of Technology Indore, Simrol, MP 453552, India}

   \date{Received XXX; accepted XXX}

 
  \abstract{The Lyman-$\alpha$ forest opacity fluctuations observed from high-redshift quasar spectra have been proven to be extremely successful in order to probe the late phase of the reionization epoch. For ideal modeling of these opacity fluctuations, one of the main challenges is to satisfy the extremely high dynamic range requirements of the simulation box,  resolving the Lyman-$\alpha$ forest while probing the large cosmological scales. In this study,  we adopt an efficient approach to model Lyman-$\alpha$ opacity fluctuations in a coarse simulation volume, utilizing the semi-numerical reionization model SCRIPT (including inhomogeneous recombination and radiative feedback) integrated with a realistic photoionization background fluctuation generating model. Our model crucially incorporates ionization and temperature fluctuations, which are consistent with the reionization model. After calibrating our method with respect to high-resolution full hydrodynamic simulation, Nyx, we compared the models with available observational data at the redshift range, $z=5.0-6.1$. With a fiducial reionization model (reionization end at $z=5.8$), we demonstrated that the observed scatter in the effective optical depth can be matched reasonably well by tuning the free parameters of our model, i.e., the effective ionizing photon mean free path and mean photoionization rate. We further pursued an MCMC-based parameter space exploration, utilizing the available data to put constraints on the above free parameters. Our estimation prefers a slightly higher photoionization rate and slightly lower mean free path than the previous studies, which is also a consequence of temperature fluctuations. This study holds significant promise for efficiently extracting important physical information about the Epoch of Reionization, utilizing the wealth of available and upcoming observational data.
}
   


   \maketitle
%

\section{Introduction}
\label{sec:intro}

The Epoch of Reionization (EoR) represents an important phase in the history of the Universe, marking the time when the first stars and galaxies form and ionize neutral hydrogen atoms (HI) in the intergalactic medium (IGM) \citep[see reviews for details,][]{2001PhR...349..125B,2009CSci...97..841C,2016ARA&A..54..313M, 2018PhR...780....1D,2022arXiv220802260G,2022GReGr..54..102C}. The precise timeline and morphology of this epoch can be crucial to understand the structure formation and evolution of the universe at those early times. There exists a handful of observational probes at various wavelengths to trace the IGM during the EoR \citep{2013ASSL..396...45Z,2016ASSL..423.....M}. Among these, spectra obtained from high-redshift quasars have been proven to be a treasure trove in extracting critical information about the nature of the IGM.

Specifically, Lyman-$\alpha$ (Ly-$\alpha$) forest absorption of quasar spectra contains information about the line-of-sight distribution of neutral hydrogen atoms (HI) in the IGM, which has been widely used in the literature to study various physical aspects. For example, these include exploiting realistic numerical simulations to constrain the photoionization rate \citep{2007MNRAS.382..325B,2011MNRAS.412.1926W,2023MNRAS.525.4093G, 2024ApJ...965..134D} and tracking the temperature evolution of the IGM \citep{2019ApJ...872...13W,2020MNRAS.494.5091G}. The high quality of existing quasar spectra also allows us to directly estimate the neutral fraction of the universe with damping wing analysis \citep{2024A&A...688L..26S} and with dark pixel analysis \citep{2015MNRAS.447..499M, 2023ApJ...942...59J}. Recently, a popular scenario of late reionization end ($z_{\mathrm{end}}\sim5.3$) has come out based on the findings of large Gunn-Peterson (GP) troughs, giving rise to excess Ly-$\alpha$ forest opacity fluctuations at $z \ge 5.3$  \citep{2021ApJ...923..223Z,2022ApJ...932...76Z,2022MNRAS.514...55B}.

Several traditional models have been found to be inconsistent in order to explain the large GP trough and the excess Ly-$\alpha$ opacity fluctuations. There exist a few suggestions to match the data, keeping the reionization end at $z\gtrsim 6$. One of these includes a hidden abundance of faint quasar populations at $z\ge5$ \citep{2020MNRAS.491.4884M}; however, it is not yet completely understood whether they have the required photon budget to make a significant contribution. An extended and hot reionization scenario can also produce significant fluctuations via the temperature dependence of recombination rate \citep{2015ApJ...813L..38D}. But, this model associates the opaque trough with large-scale galaxy overdensity \citep{2018ApJ...860..155D}, which is disfavoured by the galaxy survey results \citep{2018ApJ...863...92B, 2021ApJ...923...87C}. It is also possible that the excess ionizing background fluctuations are amplified due to very short mean free path of the ionizing photons in the IGM \citep{2016MNRAS.460.1328D, 2018MNRAS.473..560D}. However, that doesn't agree with the extrapolation from existing lower redshift measurements \citep{2014MNRAS.445.1745W}. Another emerging explanation involves a late reionization end ($z_{\mathrm{end}}\leq5.5$) where the shadowing effect on the large-scale ionization fluctuations by the residual neutral islands is responsible for the excess opacity fluctuations \citep{2019MNRAS.485L..24K, 2020MNRAS.491.1736K, 2020MNRAS.494.3080N}. However, these models are finely tuned to match the ionizing photon emissivity according to the demand of both Ly-$\alpha$ forest fluctuations and other high-redshift observables like CMB scattering optical depth. In recent times, this explanation has also been adopted by semi-numerical models, which are further able to provide constraints on the global ionization fraction utilizing Ly-$\alpha$ forest observations \citep{2021MNRAS.501.5782C, 2021MNRAS.501.4748Q,2025PASA...42...49Q}. However, these models lack a realistic description of photoionization background fluctuations and temperature fluctuations. Despite recent improvements along these lines \citep{2025arXiv250403384C}, those models are still devoid of proper calibration for individual components of opacity fluctuations against a realistic forest model. For instance, \citet{2025PASA...42...49Q} used a single simulation suite at a fixed redshift ($z=5.0$) to map out an analytic relation between true opacity and FGPA approximation via probability density function. However, as we'll discuss in our study, the FGPA relation itself may need to be modified while dealing with coarse resolution simulation, accounting for the calibration of temperature, UVB fluctuations, and redshift dependence of Ly-$\alpha$ optical depth.

Nevertheless, it is crucial to determine the strength of photoionization background fluctuations as it provides an important boundary limit to the reionization scenario. It has been shown that the fluctuations rise rapidly when the ionization bubbles start to overlap with each other, eliminating the neutral hydrogen patches \citep{2007MNRAS.382..325B,2011MNRAS.412.1926W,2018ApJ...855..106D}. The measurements of the photoionization rate ($\Gamma_{\mathrm{HI}}$) also show an increment in the redshift range $z\sim 5-6$ supporting the above statement. However, the constraints are not very stringent due to a lack of transmission flux at these high redshifts and a degeneracy with the thermal state of the IGM. In recent studies, the constraints have become more stringent with the abundance of high-quality data and robust modelling \citep{2023MNRAS.525.4093G, 2024ApJ...965..134D}.

Besides the photoionization rate, one needs the information on the photon mean free path in order to estimate the reionization source emissivities. At relatively lower redshifts, this can be measured by identifying the absorption lines corresponding to the neutral hydrogen structures \citep{2014MNRAS.438..476P}. However, this identification becomes difficult towards higher redshifts, which demands the stacking of the quasar spectra. Recently, relatively short values for the mean free path at $z\sim6$ have been reported by utilizing this stacking method \citep{2021MNRAS.508.1853B, 2023ApJ...955..115Z}, which again suggests the scenario of late reionization or excess ionization background fluctuations.

From the discussion so far, it is apparent that we need a reasonably accurate and efficient model for estimating the Ly-$\alpha$ opacity fluctuations during later stages of reionization to properly interpret the observations. However, the modelling faces the challenge of high dynamic range requirements for the simulation box. On the one hand, it has to resolve small scales ($\lesssim 100~\mathrm{Kpc}$) for a converged description of Ly-$\alpha$ forest spectra, while on the other hand, it should have a large enough scale to access the observed fluctuation spanning a wide cosmological scale. For example, \citet{2024ApJ...965..134D} utilizes a very high resolution hydrodynamic simulation, sufficient to resolve small-scale structures while their large-scale fields are not fully self-consistent. In another similar study by \citet{2023MNRAS.525.4093G}, an efficient algorithm has been exploited with a relatively larger simulation box ($160 ~h^{-1}\mathrm{cMpc}$) and coarser resolution. In principle, there may exist fluctuation features having a length scale larger than these.

In this work, we presented an efficient semi-numerical model for estimating the large-scale Ly-$\alpha$ opacity fluctuations at late stages of reionization. The motivation behind this model is to provide a reasonably accurate description of forest fluctuations within a large cosmological simulation volume while keeping the small-scale information intact. We took into account these small-scale contributions by properly calibrating the model with respect to a full hydrodynamical simulation. As we shall discuss, the model incorporated other relevant ingredients for large-scale opacity fluctuations in a self-consistent way. Further, the model also allowed us to provide constraints on background mean photoionization rate and an effective mean free path after comparing with the recent Ly-$\alpha$ opacity measurements \citep{2022MNRAS.514...55B}. The method laid out here is a proof-of-concept study with the aim of efficient EoR inference in the future.

 The layout of the paper is organized as follows: in Section \ref{sec:formalism}, we outline the formalism of our modeling along with the calibration procedure. Next, we discuss the basic properties of our model with different variants of model parameters in section \ref{sec:fid_model}. This is followed by a briefing on available observables and likelihood analysis techniques utilizing those in Section \ref{sec:likelihood}. Then in Section  \ref{sec:results},  we lay out our main findings and corresponding interpretation from the parameter space exploration study. Finally, we conclude the work in Section \ref{sec:summary}. In this paper, the default assumed cosmological parameters are $\Omega_M$ = 0.308, $\Omega_{\Lambda}$ = 0.691 $\Omega_b$ = 0.0482, $h$ = 0.678, $\sigma_8$ = 0.829 and $n_s$ = 0.961 \citep{2016A&A...594A..13P}. We will mainly work with $h^{-1}\mathrm{cMpc}$ as the distance unit unless otherwise stated.


\section{Formalism}
\label{sec:formalism}
In this section, we discuss the methodology for simulating the Ly-$\alpha$ forest optical depth distributions. This includes the calibration of the model and description of the modelling ingredients.
\subsection{Computing Ly-$\alpha$ optical depth }
In a semi-numerical set up \citep[for examples, ][]{2025PASA...42...49Q, 2025arXiv250403384C}, the optical depth for Ly-$\alpha$ absorption is generally obtained by Fluctuating Gunn Peterson Approximation \citep[FGPA;][]{1998ASPC..148...21W}. The approximation provides a relation between the Ly-$\alpha$ optical depth ($\tau_{\alpha}$) and the underlying density field ($\Delta$) of the IGM i.e. 
\begin{equation}\label{eq:tau}
    \tau_{\alpha} \propto  \Delta^{2}~T^{-0.724}~\Gamma_{\mathrm{HI}}^{-1}~(1+z)^6~H^{-1}(z)
\end{equation}
where $T$ and $\Gamma_{\mathrm{HI}}$ are the temperature and photoionization rate with $H(z)$ being the Hubble parameter at redshift $z$.
However, this relation is applicable when we can access reasonably small scales ($\sim 10-20~\mathrm{Kpc}$) structures with a very high resolution simulation. To apply this on coarse resolution semi-numerical simulations, we need to modify the relation after properly calibrating with respect to the high resolution full hydrodynamical simulation. With the aim of calibration, we assumed that for each coarse cell ($i$, with $4h^{-1}\mathrm{cMpc}$ resolution in this study), the Ly-$\alpha$ optical depth, $\tau_{i}$, follows power law relations with different physical quantities at the scale of pixel resolution. These power law indices are then calibrated with the post-processing estimates of high-resolution hydrodynamical simulations. Hence, the modified optical depth for a cell in a fully ionized state (i.e, obeying ionizing equilibrium criterion) can be written as
\begin{equation}\label{eq:tau_i}
    \tau_{i} = 5.8 \kappa \times \Delta^{2\alpha}_i\left(\frac{T_i}{7500}\right)^{-0.724\beta}\left(\frac{2.5\times 10^{-13}}{\Gamma_{\mathrm{HI},i}}\right)^{\delta}\left(\frac{1+z}{7}\right)^{6\xi-1.5}
\end{equation}
where $\Delta_i$, $T_i$ and $\Gamma_{\mathrm{HI},i}$ are the overdensity, temperature and photoionization rate for the $i$-th cell respectively with $k$ is a normalization factor. We utilized the full hydrodynamical simulation, Nyx, to estimate the different indices, as we shall discuss in the next paragraph. On the other side, the optical depth for a neutral region can be assumed to be very large, providing essentially zero transmitted flux.
Hence, the effective transmitted flux in each cell is given by 
\begin{equation}
    F_i = x_{\mathrm{HII},i}e^{-\tau_{i}} + x_{\mathrm{HI},i}e^{-\tau_{i}^n} \simeq x_{\mathrm{HII},i}e^{-\tau_{i}}
\end{equation}
where $\tau_{i}^n$ is the optical depth for the neutral region, which is sufficiently large and can be approximated to $\infty$. The ionization fraction of $i$-th cell is denoted as $x_{\mathrm{HII},i}=1-x_{\mathrm{HI},i}$. The effective optical depth of the box was then computed as
\begin{equation}
    \tau_{\mathrm{eff}} = \left \langle -\ln \left(\frac{\sum_{i=1}^{n}F_i}{n}\right)\right\rangle
\end{equation}
where $n$ is the number of pixels in a single sightline to be averaged over according to the data, and $\langle ...\rangle$ represents the averaging over all the sightlines. In this study, the averaging length scale was chosen as $\Delta z=0.1$ to be consistent with  Ly-$\alpha$ forest data \citep{2022MNRAS.514...55B} prepared from the extended XQR-30 sample.
\begin{figure}
    \centering
    \includegraphics[width=\columnwidth]{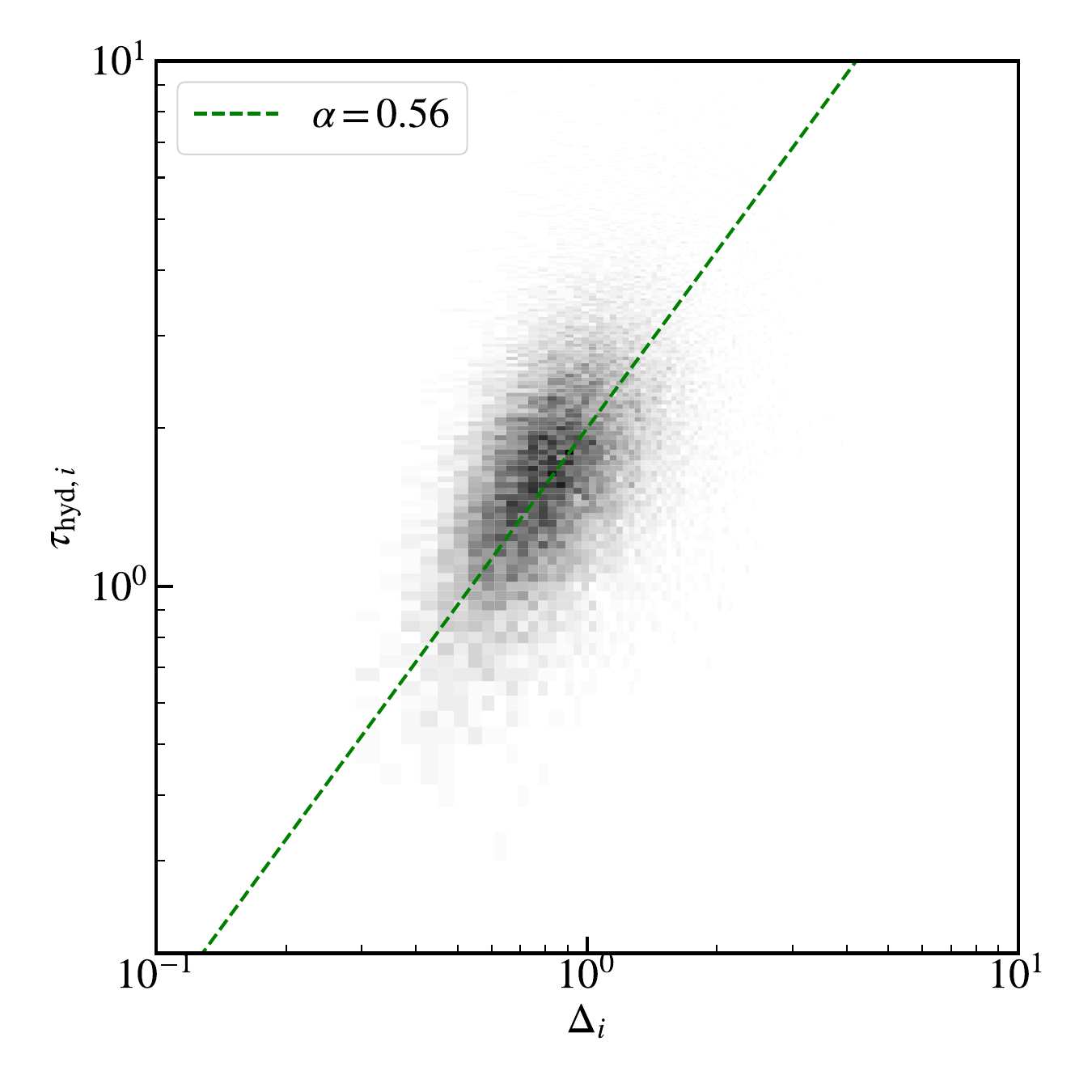}
    \caption{The figure shows the 2D density plot of optical depth ($\tau_{\mathrm{hyd},i}$) vs density ($\Delta_i$) averaged at the pixel scale ($4~h^{-1}\mathrm{cMpc}$) from NyX hydro simulation suite at redshift $z=5.0$. The slope of the correlation matches with $\alpha=0.56$, which is used to calibrate our semi-analytic model.}
    \label{fig:fig1_tau_Delta_skew}
\end{figure}
\begin{figure*}
    \sidecaption
    \centering
    \includegraphics[width=0.7\textwidth]{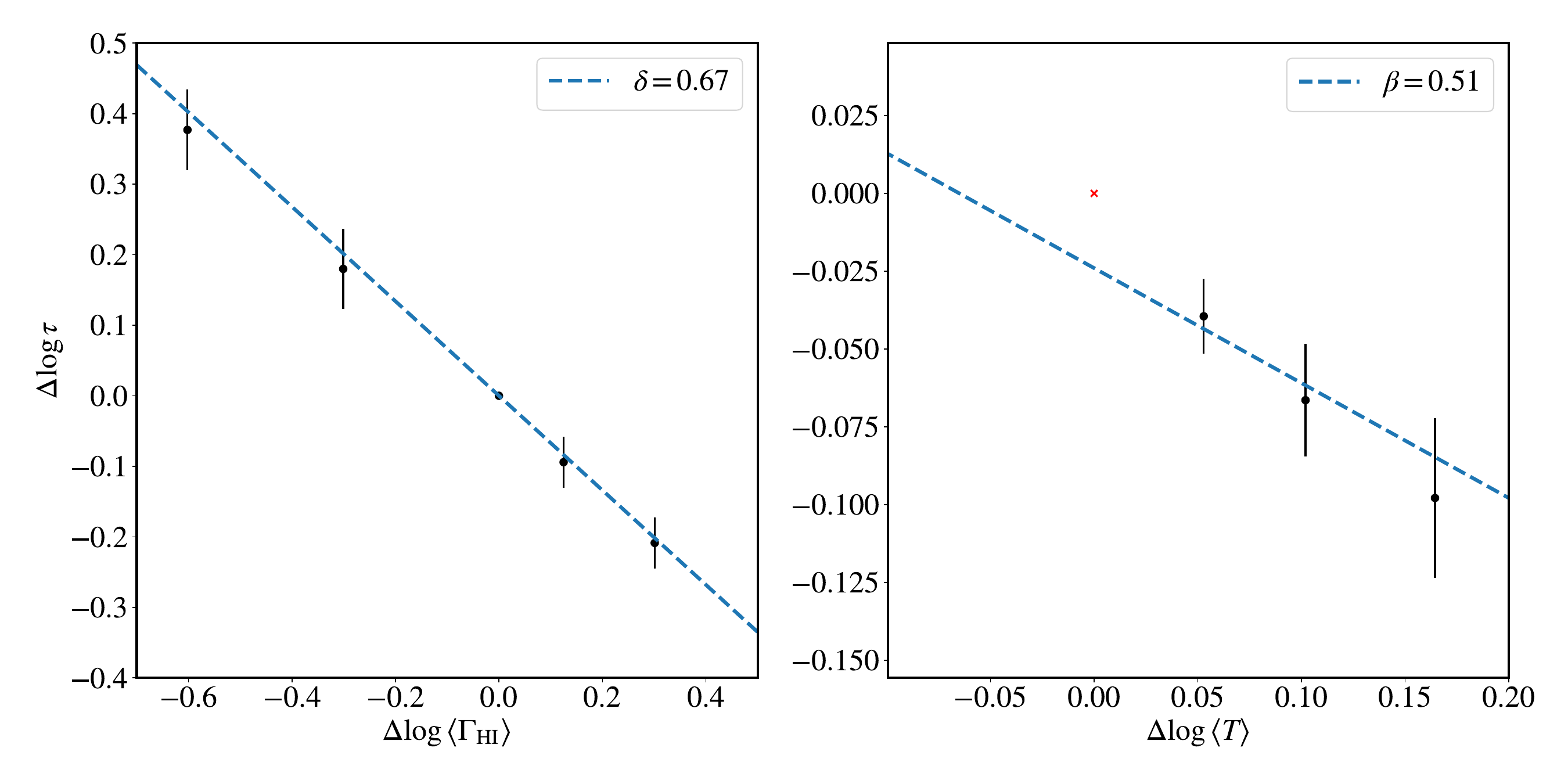}
    \caption{Global variation of optical depth ($\Delta \log\tau$) at the resolution scale ($4 h^{-1}\mathrm{cMpc}$) for variations in mean photoionization rates ($\Delta\log\langle \Gamma_{\mathrm{HI}}\rangle$, in \textit{left} panel) and for variations in temperatures ($\Delta \log \langle T\rangle$, in \textit{right} panel).}
    \label{fig:fig2_gamma_temp}
\end{figure*}
\begin{figure*}
    \centering
    \includegraphics[width=\textwidth]{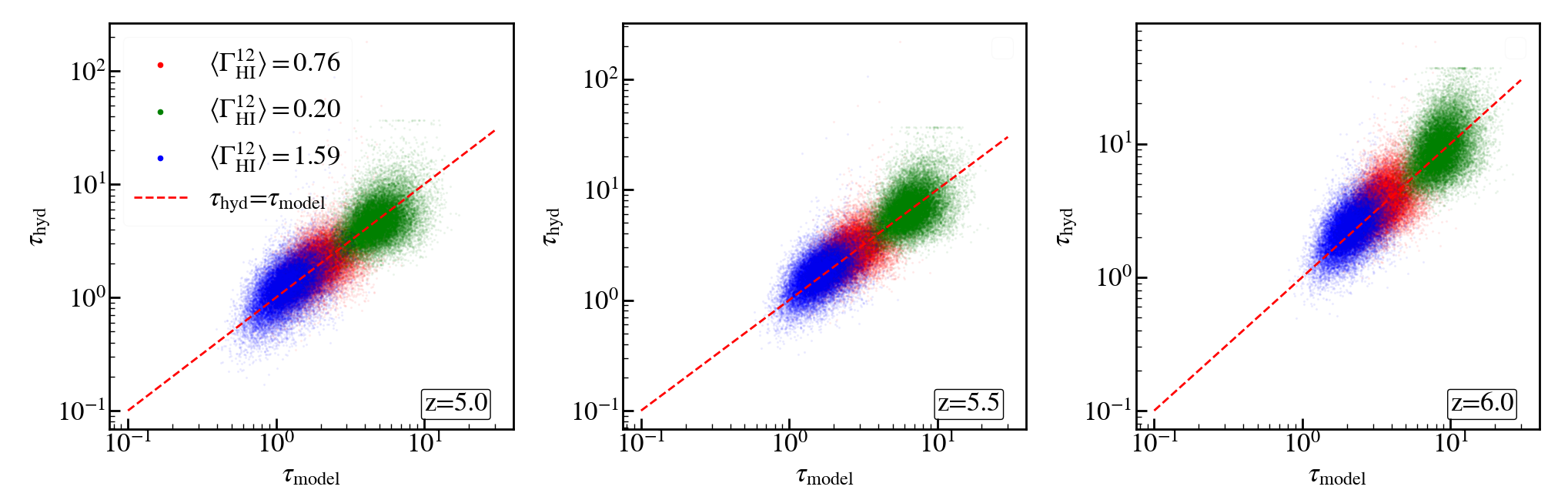}
    \caption{The comparison of optical depths from hydro simulation ($\tau_{\mathrm{hyd}}$) and calibrated empirical model ($\tau_{\mathrm{model}}$) for three different redshifts ($z=5.0,5.5$ \& $6.0$) used for calibration. The different colors indicate the set of different values of mean photoionization rates.}
    \label{fig:fig3_tau_hyd_model}
\end{figure*}

\subsection{Calibration with Nyx simulation suite}
As mentioned earlier, we utilized a cosmological hydrodynamical simulation, Nyx, to extract the skewers and calibrate our technique. The simulation description closely follows the previous work by \citet{2024ApJ...965..134D}, which provides sufficient box size ($100~h^{-1}\mathrm{Mpc}$) and resolution ($4096^3$ dark matter and baryon grids) for converged Ly-$\alpha$ forest statistics. Further, the Ly-$\alpha$ forest opacities 
are deposited in redshift space using an approximate form for the Voigt profile
\citep{2006MNRAS.369.2025T} in 2 km/s grid resolution, given the neutral hydrogen field estimated assuming ionization equilibrium. We used the skewers from this simulation suite at three redshifts ($z=5.0, 5.5$ \& $6.0$) to calibrate the redshift dependence of our empirical model. For each of these redshifts, we extracted 10,000 random skewers of overdensity and temperature fields from the simulation box. Similarly, we utilized different temperature variants of the simulation suites. The optical depth skewers also include peculiar velocity information, although the effect is supposed to be small, and we did not explicitly take into account that effect in the semi-numerical setup described later. However, due to the large-scale coherence of the velocity field, this effect can be worthwhile to explore in a future study. As this simulation assumes instantaneous reionization, these temperature variations can be obtained by changing the reionization end. We utilized the models with sudden reionization end at $z=6.2, ~6.7, ~7.7$ \& $15$; providing a range of temperature variations where earlier reionization end corresponds to lower temperature and vice versa. We also created a few different variants of effective mean photoionization rate as well by scaling the optical depth skewers in post-processing, where the opacity is inversely related to the photoionization rate.

We then binned the skewers in 3D at the scale of $\Delta x =4~h^{-1}\mathrm{cMpc}$, which is suitable for calibrating our empirical model, applicable to implement on efficient semi-numerical techniques. In Figure \ref{fig:fig1_tau_Delta_skew}, we show the distribution of optical depth skewers ($\tau_i$) with density ($\Delta_i$) at redshift, $z=5.0$ for the lowest temperature model. We found a power law correlation between them, which is consistent with $\alpha =0.56$, which corresponds to a slope of $2\alpha=1.12$ in the modified FGPA relation (Eq. \ref{eq:tau_i}). We also checked that this slope remains consistent across different redshifts and temperatures. Hence, we fixed this slope in the modified FGPA relation for further calibration. In Figure \ref{fig:fig2_gamma_temp}, we show the variations of globally averaged optical depth with respect to the mean photoionization rate ($\langle \Gamma_{\mathrm{HI}}\rangle$, at \textit{left}) and mean temperature ($\langle T\rangle$, at \textit{right}) for the variants of our simulation suite. Here, the points are the mean differences in optical depth distribution ($\Delta \log\tau$) over all the binned skewers corresponding to the differences in variants of mean photoionization rate ($\Delta\log\langle \Gamma_{\mathrm{HI}}\rangle$) and mean temperature ($\Delta\log \langle T\rangle$) while the errorbars signify 1$\sigma$ variation in the differences of optical depth distribution among skewers. We again found that the trends are consistent with the power law relation. For $\langle \Gamma_{\mathrm{HI}}\rangle$, the relation has a typical slope of $\delta=0.67$, while the temperature variations provide a typical slope of $\beta=0.51$ (excluding the point marked in red, corresponding to extreme reionization model ending at $z=15$, as the temperature relation may not be straightforward for those). As our simulation doesn’t include any heating effects (such as X-ray, Ly$\alpha$) relevant for the cosmic dawn, the temperature estimates from $z_{\mathrm{end}}=15$ model may be unrealistic. Further, this would set adiabatic cooling as the dominant mechanism at a much earlier redshift, providing an unusually cold IGM at $z<10$. This cold scenario may already be in tension with the lower redshift IGM temperature estimates (at $z=5-6$) using Ly$\alpha$ forest spike statistics \citep{2020MNRAS.494.5091G}. The model also provides a high power law index ($\gamma$) in temperature-density ($T=T_0\Delta^{\gamma-1}$) relation (around $\gamma \sim1.6$), which is again strongly disfavoured by inferred estimates using observations [see Figure \ref{fig:fig5_fid_model} for reference]. Hence, we exclude that point to keep our conclusion robust against the current observational state. We fixed the $\beta$ index to the above-mentioned value for further calibration at the pixel level. Next, we estimated the other three parameters ($\kappa$, $\delta$ \& $\xi$) by fitting the optical depth ($\tau_i$) relation for each pixel using all the simulation suites with variants of redshifts and mean photoionization rate. The bestfit slope $\delta$ is again consistent with 0.67, while the bestfit $\kappa$ and $\xi$ values are 1.28 and 0.86, respectively. In Figure \ref{fig:fig3_tau_hyd_model}, we show the comparison of optical depth distribution from the Nyx simulation and the calibrated model. We found that those are nicely correlated with each other for different redshifts (each panel) and photoionization rates (each color). This gave the confidence to use the relation within a semi-numerical setup. Notably, there is a significant scatter in the relation, which could be included \citep[e.g.][]{2025PASA...42...49Q}. However, we got satisfactory results even without that extra scatter, as we shall discuss in the next section.

\subsection{Semi-numerical setup}
In this subsection, we briefly describe semi-numerical model setup for computing the different ingredient fields, i.e., density, temperature, ultraviolet background (UVB) / photoionization, for estimating the effective optical depths, which can be further compared against the observational data. 

\subsubsection{Generating density and collapsed halo field}
For generating the cosmological density field, we used the approach adopted in the publicly available semi-numerical reionization model 21cmFAST \citep{2011MNRAS.411..955M}. The initial conditions on density and velocity fields are generated in Lagrangian space following the gravitational evolution of the fields via Zel'Dovich Approximation \citep{1970A&A.....5...84Z}. This approximation has been proven to be successful in closely resembling the full N-body simulations with less computational cost. We used a simulation of box size $256~h^{-1}\mathrm{cMpc}$, which is sufficiently large to serve the purpose of this study. After having the particle fields, we gridded those at a high resolution ($0.5 h^{-1}\mathrm{cMpc}$) and then averaged them to $\Delta x=4h^{-1}\mathrm{cMpc}$, which is the resolution we adopted throughout the study. 

Besides, to construct the collapsed halo mass fraction, we adopted the Excursion Set Formalism (ESF-L) in Lagrangian space, following another semi-numerical reionization model AMBER \citep{2022ApJ...927..186T}. To be specific, the formalism uses Extended Press Schecter theory, which computes the collapsed mass fraction ($f_{\mathrm{coll}}$) above a minimum threshold halo mass ($M_{\mathrm{min}}$) given the filtered overdensity at a cosmological scale. Here, we used a spherical tophat filter for filtering the overdensity at a certain scale ($2~\mathrm{cMpc}$), which provides a good match with the full N-body simulation. At each redshift, we generated a collapsed fraction field for a logarithmic minimum mass ($\log M_{\mathrm{min}}$) interval of $0.1$ in the range $7-12$. 

\begin{figure*}
    \centering
    \includegraphics[width=\textwidth]{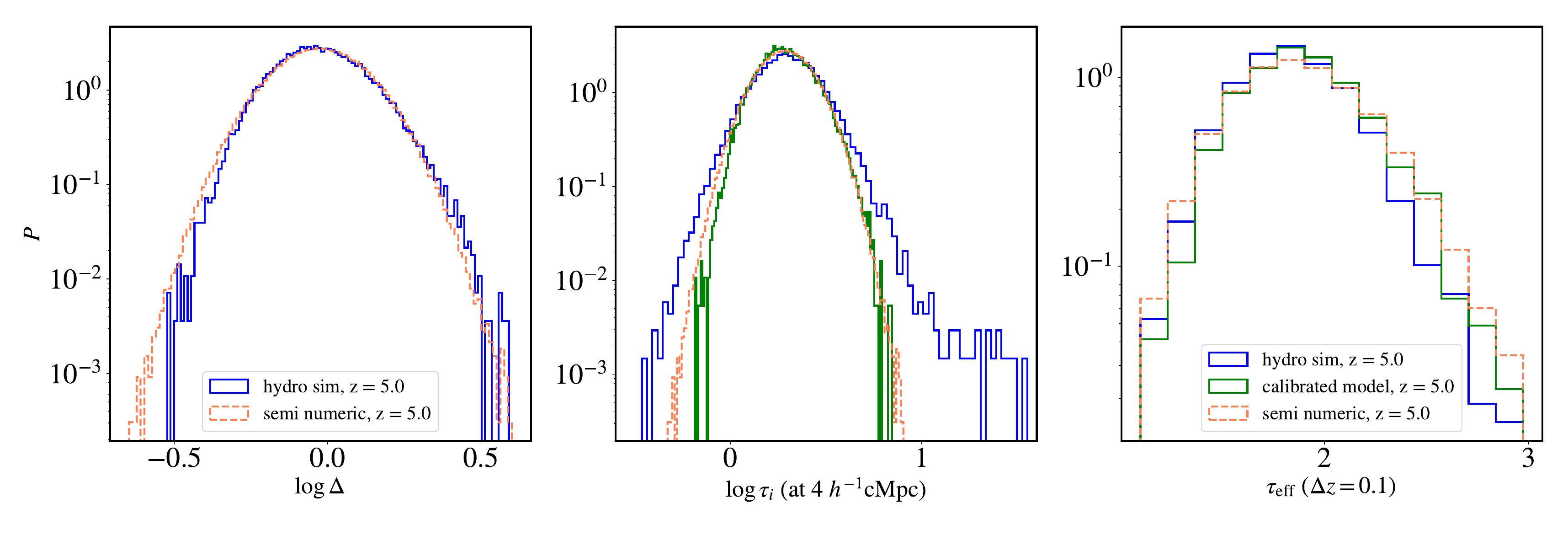}
    \caption{\textit{Left Panel:} The distribution of densities ($\log \Delta$) from the hydro simulation (blue) and the semi-numerical setup (dashed orange). \textit{Middle Panel:} The distribution of optical depths averaged at $4h^{-1}\mathrm{cMpc}$ scale ($\tau_i$) for hydro simulation (blue), calibrated empirical relation (green), and calibrated semi-numerical setup (dashed orange). \textit{Right Panel:} The distributions of effective optical depth skewers ($\tau_{\mathrm{eff}}$) for the previous three cases.}
    \label{fig:fig4_tau_delta_dist}
\end{figure*}

In the left panel of Figure \ref{fig:fig4_tau_delta_dist}, we show the overdensity distribution of the Nyx simulation (\textit{blue}) and the generated overdensities using semi-numerical method (\textit{dashed orange}) as discussed above. We checked that the natures of these two distributions are in agreement with each other. This resemblance is useful to avoid any discrepancies in optical depth due to the background density field. 

\subsubsection{Semi-numerical reionization model}
We utilized an explicitly photon-conserving semi-numerical reionization model, \texttt{SCRIPT}, which provides the ionization and thermal state of the universe in a cosmologically representative simulation volume \citep{2018MNRAS.481.3821C, 2022MNRAS.511.2239M}. To initiate the model, we provided inputs comprising the density field and the distribution of collapsed halos capable of emitting ionizing radiation as generated by the method discussed earlier. For our model to compute the complete reionization history, we utilized comoving simulation boxes spanning from $z = 15$ to $z = 5$, with an interval of $\Delta z=0.1$. However, for the purpose of this study, we essentially focused on the redshift range $z=5-6.1$.

Our model utilizes the photon-conserving algorithm to construct the reionization topology within the simulation box. The ionization field relies on the ionization efficiency parameter $\zeta(M_h, z)$, which estimates the available ionizing photons per hydrogen atom. This parameter may vary based on halo mass ($M_h$) and redshift ($z$). To address inhomogeneous recombinations, we adjust the ionization criteria to compensate for excess neutral atoms, while introducing small-scale fluctuations through a globally averaged clumping factor, $C_{\mathrm{HII}}$. This parameter remains highly uncertain due to the absence of direct observational measurements. Recently, a claim of a relatively high clumping factor ($\sim12$) has been made via an analytical argument applied to photoionization rate and mean free path measurements
\citep{2024arXiv240618186D}, although most simulation studies suggest a value around $2-5$ \citep[e.g.][]{2020ApJ...898..149D,2025MNRAS.539L..18A}.
In this study, we consistently fixed this parameter at a value of 3. To calculate the recombination number density, it is necessary to consider the thermal history of the medium. Therefore, alongside tracking the ionization process, we also simulated the temperature evolution within each grid cell of our computational volume. The code is designed to automatically factor in the effects of spatially inhomogeneous reionization on temperature evolution. Specifically, when a region becomes ionized for the first time, its temperature is assumed to increase by a specific amount, denoted as  $T_{\mathrm{re}}$ \citep{1997MNRAS.292...27H,2009ApJ...701...94F,2018MNRAS.477.5501K,2022MNRAS.511.2239M}.  

\begin{figure*}
    \centering
    \includegraphics[width=\textwidth]{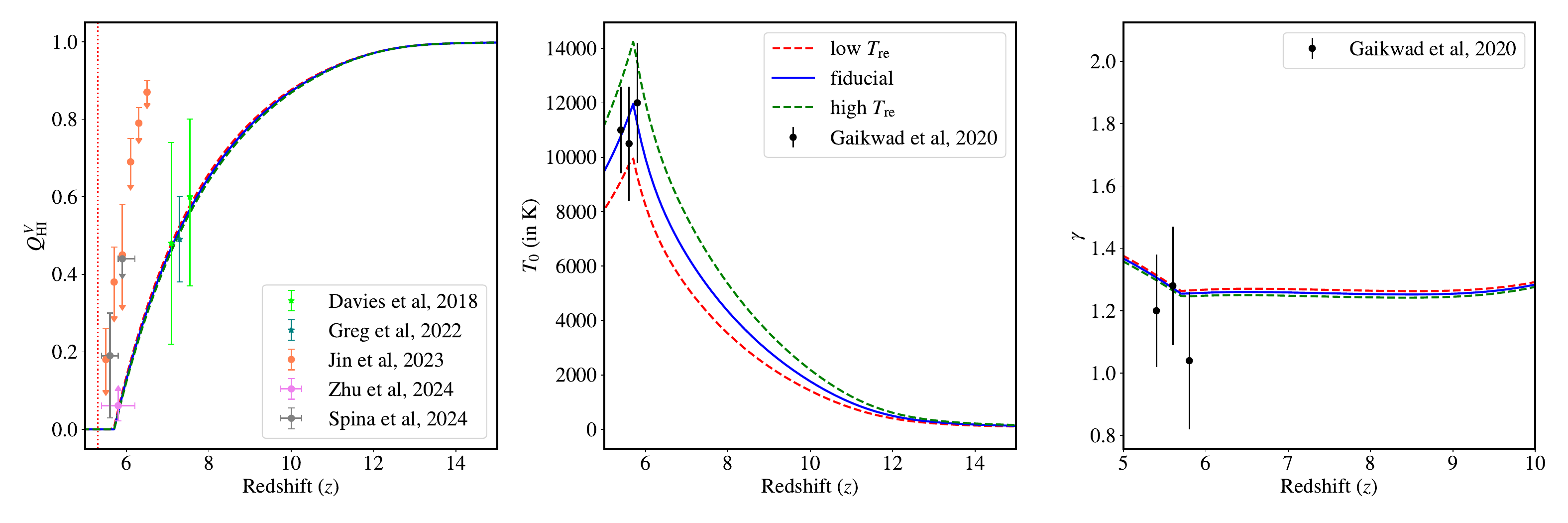}
    \caption{The figure shows the neutral fraction ($Q_{\mathrm{HI}}^V$), mean IGM temperature ($T_0$), and $T-\Delta$ power law index ($\gamma$) evolution with redshift for our fiducial reionization model (\textit{blue solid} lines) along with other two variants using different reionization temperature increment, $T_{\mathrm{re}}$ (high: \textit{green dashed}; low: \textit{red dashed}). We also show the recent observational constraints on the neutral fractions \citep{2018ApJ...864..142D,2022MNRAS.512.5390G,2024MNRAS.533L..49Z,2024A&A...688L..26S} and IGM temperatures \citep{2020MNRAS.494.5091G}.}
    \label{fig:fig5_fid_model}
\end{figure*}
Additionally, our model accounts for radiative feedback, which inhibits the production of ionizing photons in halos that experience gas heating. We followed the `step feedback' framework introduced in \citet{2022MNRAS.511.2239M}, where halos below a certain mass threshold, defined as $M_{\mathrm{min}} = \mathrm{Max} \left[M_{\mathrm{cool}}, M_{J}\right]$, are assumed to retain no gas, while those above retain all their gas. Here, $M_{\mathrm{cool}}$ is the atomic cooling threshold mass, and $M_J$ is the Jeans mass at virial overdensity. This simplified step-function approach to radiative feedback is both efficient and effective for modeling reionization. Importantly, $M_{\mathrm{min}}$ varies with redshift and across spatial locations, especially in ionized zones where temperature-driven feedback becomes significant. The Jeans mass, being temperature-dependent, contributes to this variability. In contrast, the neutral regions are not impacted by feedback, where $M_{\mathrm{min}}$ aligns with $M_{\mathrm{cool}}$, which is roughly $\sim 3\times 10^7 M_{\odot}$ at $z = 20$ and $\sim 2\times 10^8 M_{\odot}$ at $z = 5$. In feedback-affected zones, $M_{\mathrm{min}}$ typically surpasses $10^9 M_{\odot}$.
.
\subsubsection{Model for UVB fluctuations}
The fluctuations in HI photoionization rate are a crucial component in determining the Lyman-$\alpha$ optical depth during the later stages of the reionization era. These mainly arise due to the discrete nature and clustering of ionizing sources, which remains even after ionization equilibrium is established in the post-ionization phase \citep{2009MNRAS.400.1461M}.  However, this is hard to accurately model within the standard semi-numerical reionization framework due to the lack of small-scale information.

In this work, we deployed the model developed by \citet{2016MNRAS.460.1328D}, which incorporates the effect of fluctuating mean free path on the photoionization rate. The photoionization rate in each cell ($i$) is computed by integrating the product of specific ionizing intensity (including local, $J_{\nu,i}^\mathrm{L}$ and non local, $J_{\nu,i}^{\mathrm{NL}}$ contribution) and interaction cross section ($\sigma_{\mathrm{HI}}$) across UVB photon frequency ($\nu$), i.e.
\begin{equation}\label{eq:Gamma}
    \Gamma_{\mathrm{HI},i} = 4\pi \int_{\nu_{\mathrm{HI}}}^{4\nu_{\mathrm{HI}}}\left[\frac{J_{\nu,i}^\mathrm{L} + J_{\nu,i}^{\mathrm{NL}}}{h\nu}\right]\sigma_{\mathrm{HI}}(\nu)\mathrm{d}\nu
\end{equation}
This integration over the frequency makes the computation slightly inefficient, although it has been checked that a single frequency ($\nu\sim1.32\nu_{\mathrm{HI}}$) in the middle provides a good approximation \citep{2016MNRAS.460.1328D}, which is sufficient for this study. To compute this relation, we need to estimate the non-local contribution, which is given as 
\begin{equation}\label{eq:nonlocal}
    J_{\nu,i}^{\mathrm{NL}} = \sum\limits_{j \ne i}^N \frac{L_{\nu,j}}{4\pi r_{ij}^2}\exp\left[-\int_{r_j}^{r_i}\frac{\mathrm{d}x}{\lambda(x)}\right]
\end{equation}
where $r_{ij}$ is the distance between $i$-th and $j$-th cells; $\lambda(x)$ is the mean free path at each position along the sightlines, and $L_{j,\nu}$ is the specific luminosity on the corresponding ($j$-th) cell. This requires the mean free path variation for each cell that is computed as 
\begin{equation}\label{eq:lambda}
    \lambda_{\nu}(x) \propto \lambda_0~\Delta^{-1}(x)~\Gamma_{\mathrm{HI}}^{2/3}(x)~\nu^{0.9}
\end{equation}
Here, $\lambda_0$ gives a proxy of the averaged mean free path, which can be treated as a free parameter in the model, and $\Delta$ is the overdensity distribution. The dependencies on the different physical parameters via the above-mentioned power law indices come from realistic assumptions about absorbing gas density distribution; however, small variations in these indices do not change the fluctuations significantly \citep{2023MNRAS.525.4093G}. On the other hand, the local contribution takes into account the intensity budget within a cell, which is estimated using an approximation \citep{2016MNRAS.460.1328D} to get a resolution-independent average photoionization rate, i.e. 
\begin{equation}\label{eq:local}
    J_{\nu,i}^\mathrm{L} \approx \frac{\epsilon_{\nu,i}\lambda_{\nu,i}}{4\pi}\left(1-e^{-0.72\Delta x/\lambda_{\nu,i}}\right)
\end{equation}
with $\Delta x$ is the resolution of the cell and $\epsilon_{\nu,i}$ is the specific emissivity. The equations \ref{eq:Gamma} and \ref{eq:nonlocal} can be solved iteratively in each cell to get the UVB fluctuations field.

\subsubsection{Comparison with hydro skewers}
In the middle and right panels of Figure \ref{fig:fig4_tau_delta_dist}, we show the optical depth distribution ($\tau_i$) at pixel scale ($4 h^{-1}\mathrm{cMpc}$) and effective optical depth ($\tau_{\mathrm{eff}}$, averaged at $\Delta z=0.1$) respectively at a fiducial redshift, $z=5.0$. The distribution of opacities from hydro skewers is shown in \textit{blue}, the optical depths estimated using our calibrated empirical relation on hydro density and temperature field are shown in \textit{green}, while the estimates from the semi-numerical methods (the fiducial model has been discussed separately in the next section) are in \textit{dashed orange}. We found that the fluctuations at the pixel scale are slightly lower for the semi-numerical setup than the full hydro simulation, which is not very surprising due to the involvement of complex astrophysical processes in those simulations and the fact that we do not introduce the additional scatter in the modified FGPA relation. However, the $\tau_{\mathrm{eff}}$ distributions are very similar to each other, providing a check on the robustness of the calibration procedure. One possible reason behind these findings may be the contribution of large-scale correlations in peculiar velocities, which are present in the hydro simulation, which we neglect in the semi-numerical model. The peculiar velocity component is ideally a part of the FGPA relation that can suppress the scatter in optical depth while averaging along the line of sight. However, this needs to be investigated in a separate study later, after a proper comparison of velocity fields between the semi-numerical model and hydro simulation.

\begin{figure*}
    \centering
    \includegraphics[width=\textwidth]{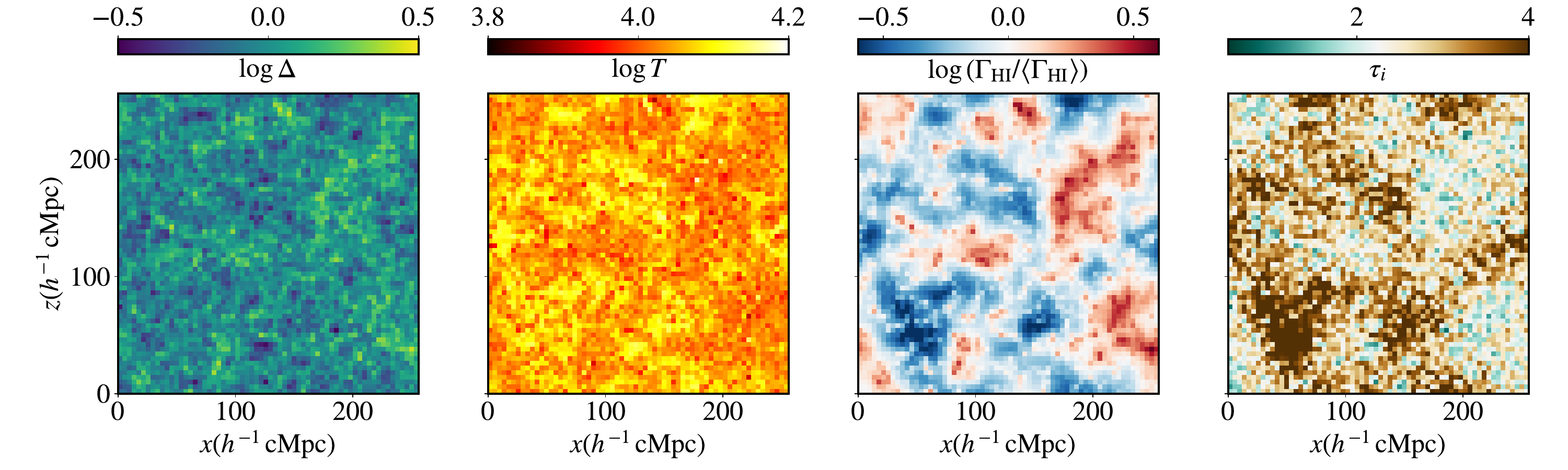}
    \caption{The snapshots of density ($\Delta$), temperature ($T$), UVB fluctuations ($\Gamma_{\mathrm{HI}}/\langle\Gamma_{\mathrm{HI}}\rangle$) and Ly-$\alpha$ optical depth ($\tau_i$), gradually from \textit{left} to \textit{right} using a fiducial setup of the calibrated semi-numerical model at redshift, $z=5.4$.}
    \label{fig:fig6_snapshots}
\end{figure*}
\begin{figure*}
    \centering
    \includegraphics[width=\textwidth]{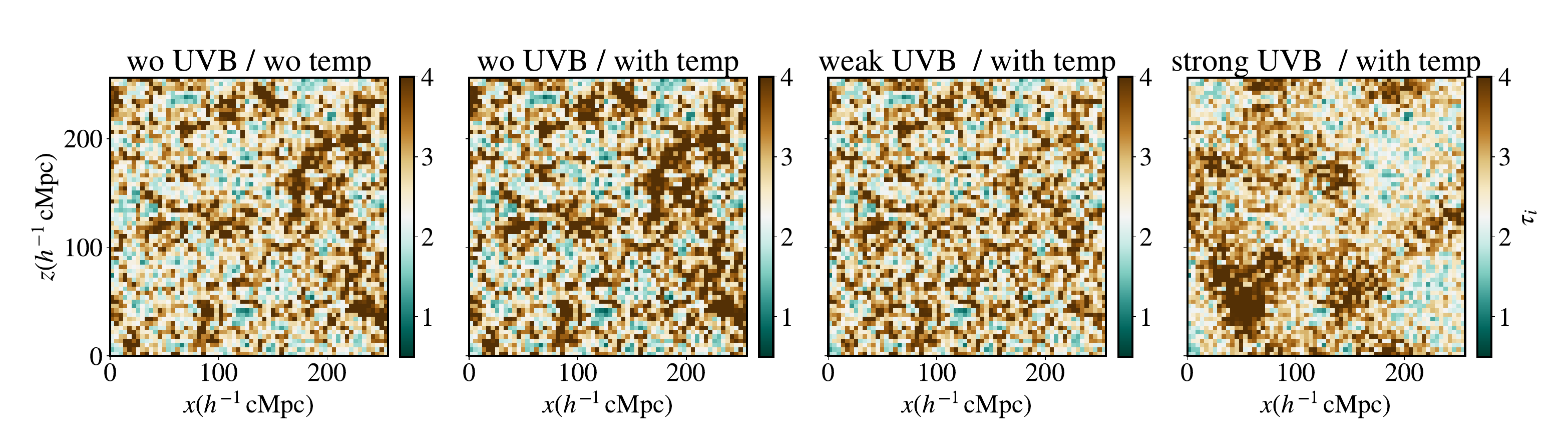}
    \caption{The snapshots of Ly-$\alpha$ optical depth for different combinations of temperature and UVB fluctuation inputs. From left to right, we show the cases gradually where we include i) no UVB/temperature fluctuation; ii) only temperature fluctuation but no UVB fluctuations; iii) weak UVB fluctuations with temperature fluctuations, and iv) strong UVB fluctuations with temperature fluctuations.}
    \label{fig:fig7_tau_snapshots}
\end{figure*}

\section{Demonstration with fiducial model}
\label{sec:fid_model}
In this section, we demonstrate the features of our model assuming a fiducial reionization model. We chose a fiducial model from the allowed posterior parameter spaces satisfying a variety of available observational constraints \citep{2022MNRAS.515..617M}, including CMB scattering optical depth, UV luminosity functions, and low-density IGM temperature estimates. In Figure \ref{fig:fig5_fid_model}, we show the evolution of neutral fractions (reionization history), mean temperature evolution, and power law index ($\gamma$) evolution of $T-\Delta$ relation for our fiducial model. We show the recent observational constraints on these quantities as well, i.e. \citep[neutral fractions;][]{2018ApJ...864..142D,2022MNRAS.512.5390G,2023ApJ...942...59J,2024MNRAS.533L..49Z,2024A&A...688L..26S} and \citep[IGM temperature; ][]{2020MNRAS.494.5091G}. It is apparent that the fiducial model obeys all the constraints reasonably well. We also show two variants of the model using different temperature increments ($T_{\mathrm{re}}$) values representing two extremes. We chose these models such that they just become consistent with the observational constraints on mean IGM temperature from the two ends. Later, we'll use these models to marginalize over different allowed temperatures while estimating the photoionization rate and mean free path.

Next, we studied the effects of varying the parameters that are responsible for generating the UVB fluctuations. 
In Figure \ref{fig:fig6_snapshots}, we show the snapshots of various physical quantities for the fiducial model at redshift $z=5.4$. These are density ($\Delta$), temperature ($T$), UVB fluctuations ($\Gamma_{\mathrm{HI}}/\langle \Gamma_{\mathrm{HI}}\rangle$) and the Ly-$\alpha$ optical depth ($\tau$) from left to right respectively. The correlation between different fields is quite apparent from the snapshots. The overdense regions are ionized at a relatively earlier time, allowing more time for cooling. Hence, the temperatures in those regions are comparatively lower. The photoionization rate, on the other hand, is larger for high density regions. This is expected as high-density regions naturally host galaxies and stars, which are supposed to be dominant sources of ionizing photons. In this fiducial scenario, we have used $\lambda_0=10~h^{-1}\mathrm{cMpc}$ for the UVB field. For this small mean free path, the opacity fluctuations are mainly dominated by the ionization background fluctuation. However, the strength of this fluctuation can vary depending on the mean free path parameter, which we discuss later.

\begin{figure*}
    \centering
    \includegraphics[width=\textwidth]{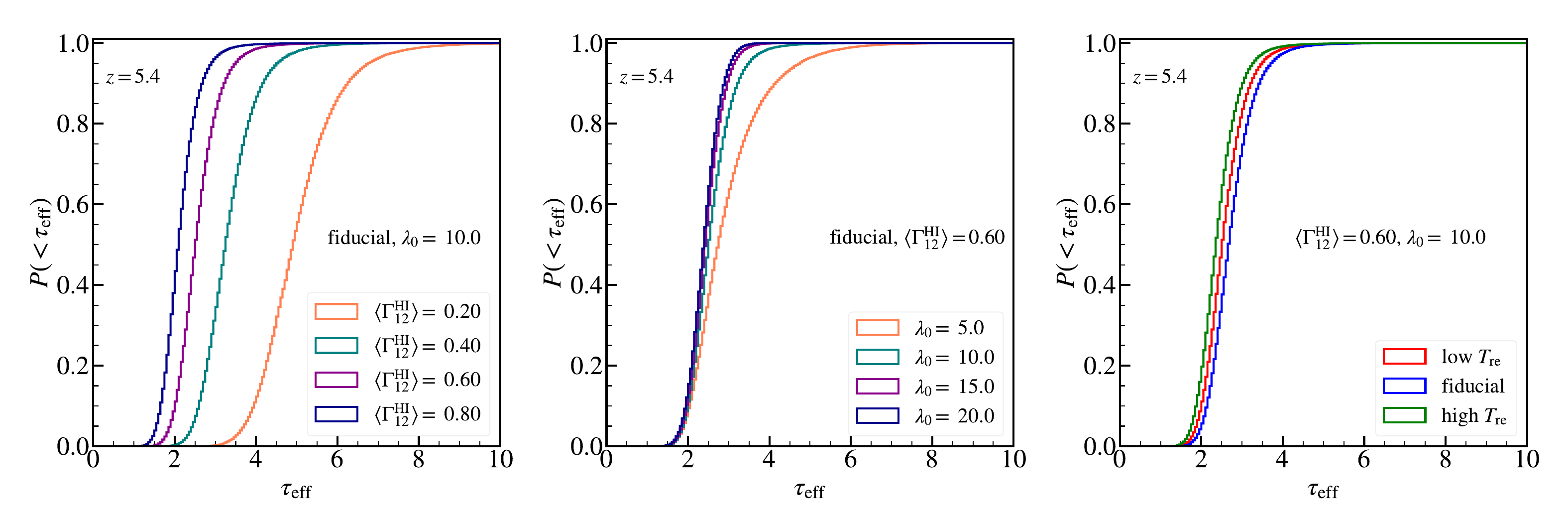}
    \caption{The figure shows the cumulative distribution (CDF) of $\tau_{\mathrm{eff}}$ at redshift, $z=5.4$ for a variety of parameter values. The \textit{left} panel shows the variation of CDF for different mean photoionization rate (in $10^{-12}/s$, i.e., $\langle\Gamma_{12}^{\mathrm{HI}}\rangle$). The \textit{middle} panel shows the CDF variation with effective mean free path ($\lambda_0$ in $h^{-1}\mathrm{cMpc}$), the \textit{right} panel shows the CDF variation with different reionization temperature increment ($T_{\mathrm{re}}$).}
    \label{fig:tau_eff_lambda_gamma}
\end{figure*}

In Figure \ref{fig:fig7_tau_snapshots}, we demonstrated the effect of UVB and temperature fluctuations on the Ly-$\alpha$ opacities. At the left panel, we show the case where no UVB or temperature fluctuations are included, and the opacities are driven solely by the density field. We included temperature fluctuations in the next panel, which increases the fluctuations as overdense regions become more opaque due to lower temperature and vice versa. On the third panel, we incorporated a weakly fluctuating UVB field, which slightly neutralises the fluctuations. This happens as the high-density regions have higher photoionization rates, contributing to lower opacities, which further results in smoothening the optical depth field. At the rightmost panel, the fluctuations are again increased due to a strong UVB field.
In Figure \ref{fig:tau_eff_lambda_gamma}, we show the cumulative probability distributions (CDF) of the effective optical depth ($\tau_{\mathrm{eff}}$) for different mean photoionization rate (in $10^{-12}/s$ i.e. $\langle\Gamma_{12}^{\mathrm{HI}}\rangle$, at left) and effective mean free path ($\lambda_0$, at right). It can be seen that the mean photoionization rate mainly shifts the mean value of $\tau_{\mathrm{eff}}$ while $\lambda_0$ contributes to the amount of fluctuation in the distribution.  A lower value of the mean photoionization rate corresponds to a higher effective optical depth distribution and vice versa. This is also expected from equation \ref{eq:tau_i}, where these two quantities are inversely related to each other. Physically, the photoionization rate also controls the transmission flux, i.e., a higher photoionization rate provides a higher transmission, which corresponds to a lower optical depth. On the other hand, the mean free path essentially provides an estimate of how far a photon can travel before getting absorbed by the residual neutral gas. Hence, a higher value of $\lambda_0$ results in a smoother transmission having less amount of fluctuations due to the absorption, which further narrows down the optical depth distribution.

 \begin{figure*}
 \sidecaption
     \centering
     \includegraphics[width=0.7\textwidth]{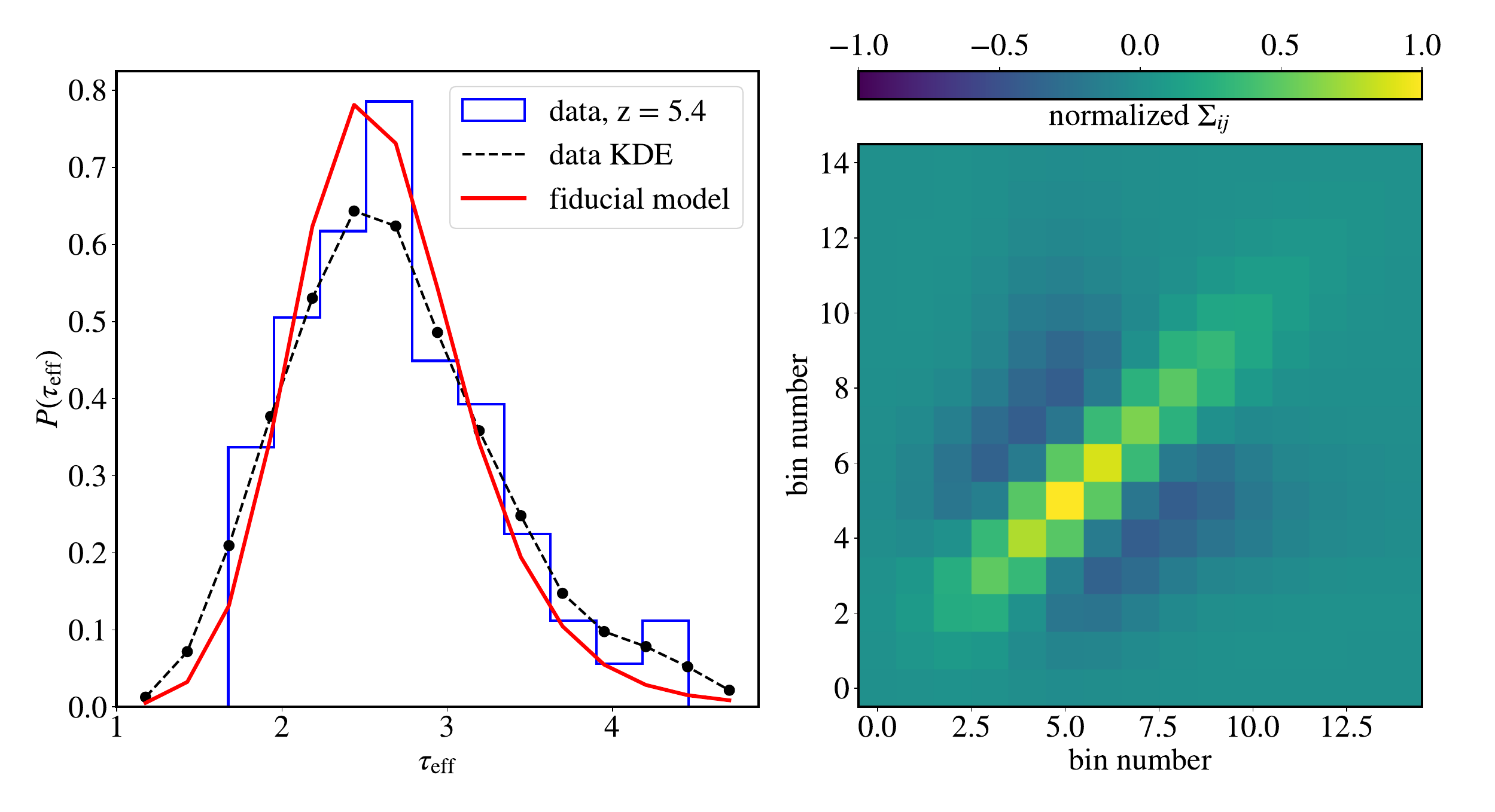}
     \caption{\textit{Left Panel:} An example of Probability Distribution Functions (PDF) from the data and the fiducial model at redshift, $z=5.4$. The blue line shows the histogram obtained from the data, while the dashed black line shows the fit using Gaussian KDE. Similarly, the red line shows the distribution obtained from our fiducial model. For the likelihood analysis, we use 15 equally spaced bins as marked by the black solid points. \textit{Right Panel:} The normalized covariance matrix at $z=5.4$.}
     \label{fig:fig8_binned_pdf}
 \end{figure*}

 \begin{figure*}
    \centering
    \includegraphics[width=\textwidth]{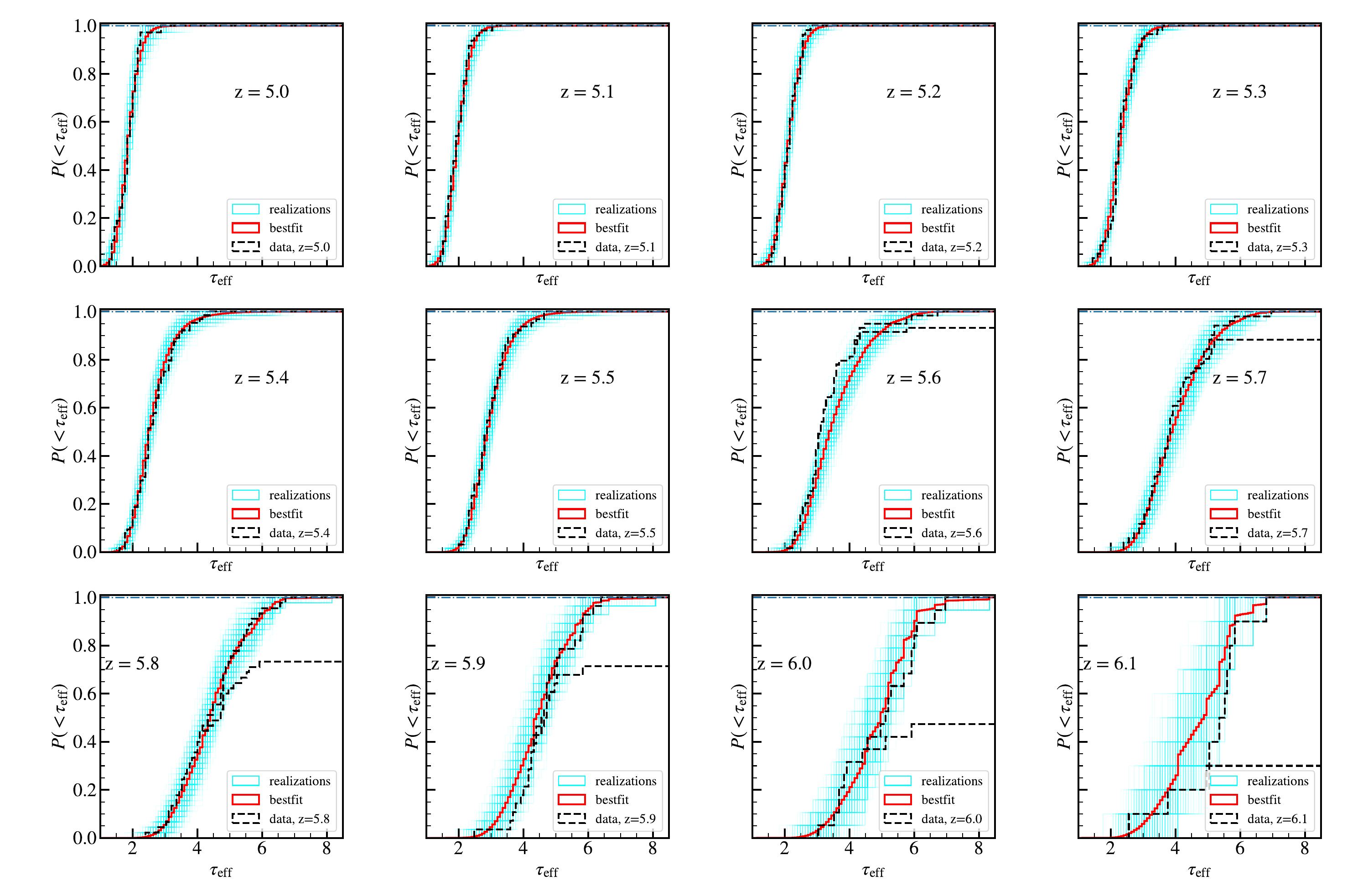}
    \caption{The CDF plots of $\tau_{\mathrm{eff}}$ for different redshifts ($z=5.0-6.1$). Each panel shows the combined CDFs of the bestfit model using 500 random realizations ($\textit{red}$) along with the individual realization ($\textit{cyan}$) and observational data ($\textit{black}$). We show both the optimistic and pessimistic scenarios for the data. }
    \label{fig:tau_eff_ionfrac}
\end{figure*}

\begin{figure*}
    \centering
    \includegraphics[width=0.95\textwidth]{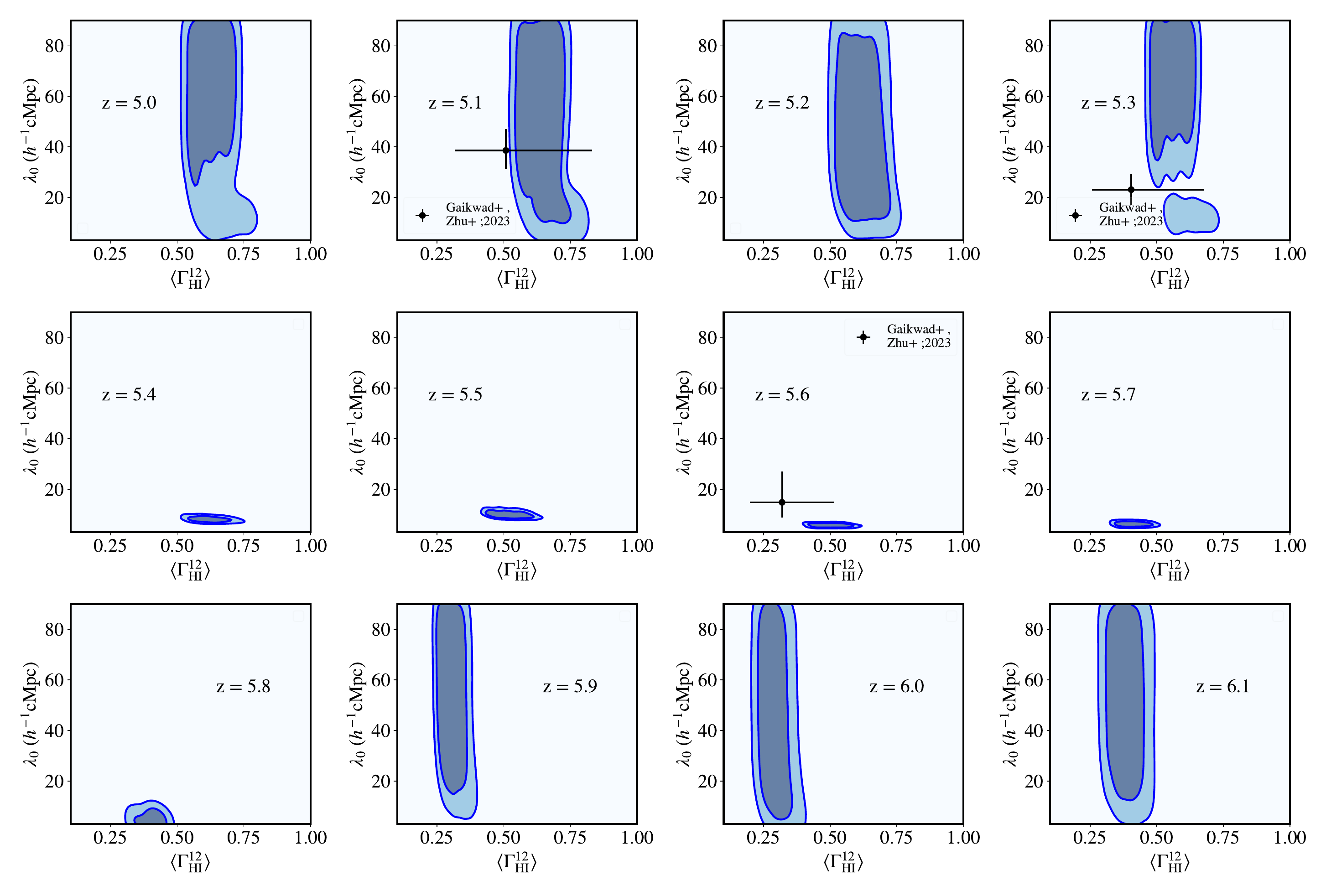}
    \caption{The joint 2D posterior distribution of effective mean free path ($\lambda_0$) and mean photoionization rate ($\langle\Gamma_{12}^{\mathrm{HI}}\rangle$)  for different redshifts (from $z=5.0$ at top left to $z=6.1$ at bottom right). }
    \label{fig:fig10_Gamma_lambda_mcmc_posterior}
\end{figure*}

 \begin{figure}
     \centering
     \includegraphics[width=0.9\columnwidth]{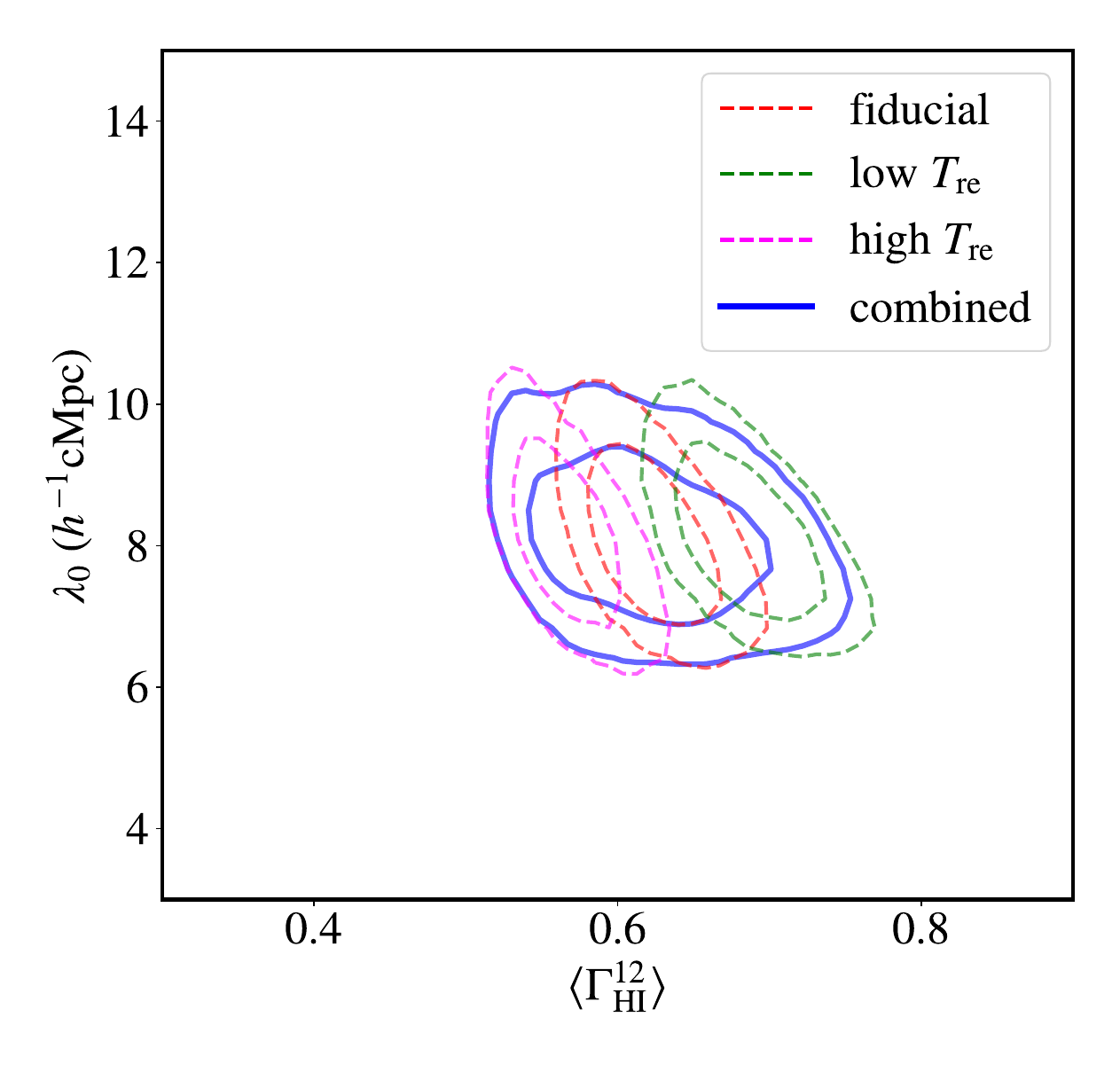}
     \caption{2D posterior distribution between $\lambda_0$ and $\Gamma_{\mathrm{HI}}$ for different temperature models separately along with the combined contours. The contours are shown at 68\% and 95\% levels.}
     \label{fig:fig11_2d_post_demons}
 \end{figure}

\section{Observational data \& likelihood analysis}
\label{sec:likelihood}
We utilized the high quality Ly-$\alpha$ forest data from 67 high-redshift quasars following \citet{2022MNRAS.514...55B}. This sample includes 51 optical quasar spectra from VLT X-shooter spectrograph, among which 25 are from the XQR-30 program \citep{2023MNRAS.523.1399D} and 26 are from archival X-shooter data. The remaining 16 spectra have been taken with the Keck ESI spectrograph. All these high redshift 
 (at a redshift $z\ge 5.6$) quasars spectra with SNR $\ge 10$ per pixel provide an ideal dataset to probe the late phase of the reionization epoch. However, the quasar continuum is very uncertain at these redshifts due to the lack of enough pixels to recover the continuum flux. Hence, to compute effective optical depth, a novel principal component analysis (PCA) method has been followed in recent studies \citep{2018ApJ...864..143D,2021MNRAS.503.2077B}. Then the effective optical depths are computed with the bins of $\Delta z=0.1$  \citep{2022MNRAS.514...55B}. 
 
 In our analysis, the observational uncertainties and wavelength-dependent continuum uncertainties are taken into account via forward modelling. To be specific, the measurement uncertainties are added with each realization of the model skewers according to the data, while the continuum uncertainties are multiplicative. The sightlines for which no significantly detected transmission is available,  are treated as optimistic limits, i.e., assumed to be $2\sigma$ where $\sigma$ is the data uncertainty corresponding to that sightline. In a pessimistic scenario, the flux could be assumed to be zero, producing an infinitely opaque sightline \citep[following, ][]{2022MNRAS.514...55B}.

 Given the above dataset, we used the standard Bayesian approach for parameter space exploration. Specifically, we computed the conditional probability distribution or the posterior $\mathcal{P}(\theta \vert \mathcal{D})$ of the model parameters $\theta$, provided the observational data sets $\mathcal{D}$ as mentioned in the previous paragraph. This can be estimated using the Bayes theorem
\be\label{eq:bayes_eq}
 \mathcal{P}(\theta\vert \mathcal{D})=\frac{\mathcal{L}(\mathcal{D} \vert \theta) ~\pi(\theta)}{\mathcal{P}(\mathcal{D})},
\ee
where $\mathcal{L}(\mathcal{D} \vert \theta)$ is the conditional probability distribution of data given the parameters or the likelihood, $\pi(\theta)$ is the prior and $\mathcal{P}(\mathcal{D})$ is the evidence (which can be treated as the normalization parameter and does not play any role in our analysis). 

However, the raw dataset can not be used for likelihood analysis due to the dominance of Poisson noise arising from its discrete nature. To this end, different summary statistics, such as Cumulative Distribution Function (CDF), mean fluctuations, etc can be used for inference studies with the assumption that the posterior $\mathcal{P}(\theta\vert \mathcal{D}) \approx \mathcal{P}(\theta\vert \mathcal{D_S})$, where $\mathcal{D_S}$ is the corresponding statistics derived from data. Among these, Probability Distribution Functions (PDFs) fitted with Gaussian Kernel Density Estimator (KDE) are commonly used \cite{2021MNRAS.506.4389G,2021MNRAS.501.5782C} for robust likelihood inference. Here, we adopted a similar approach and evaluated the KDE distribution into 15 bins to compute the likelihood.
We did the exact same procedure with both the observational data and our models. We kept the number of sightlines same as the data. Once we had the binned probability distribution, we computed the logarithm of the likelihood ($\mathcal{L}$) distribution as
 \begin{equation}
 \begin{aligned}
     -2\ln{ \mathcal{L}} &= \chi^2 \\ & = \sum\limits_{i,j}[p_{\mathcal{D}}(\tau_{\mathrm{eff}})-p_{\mathcal{M}}(\tau_{\mathrm{eff}})]_i\Sigma_{ij}^{-1}(\mathcal{M})[p_{\mathcal{D}}(\tau_{\mathrm{eff}})-p_{\mathcal{M}}(\tau_{\mathrm{eff}})]_j
 \end{aligned}
 \end{equation}
 where $p_{\mathcal{D}}(\tau_{\mathrm{eff}})$ is the observed KDE PDF of effective optical depth and $p_{\mathcal{M}}(\tau_{\mathrm{eff}})$ is the corresponding model KDE PDF. We also took into account the correlations between the bins by computing the covariance matrix $\Sigma_{ij}(\mathcal{M})$, which is model dependent. This was computed using 1000 random independent realizations of the sightlines.

 In the left panel of Figure \ref{fig:fig8_binned_pdf}, we show the probability distributions of the data along with the fiducial model at redshift, $z=5.4$. The blue histogram represents the effective optical depth distribution according to the original data, while the black dashed line is achieved after applying the Gaussian KDE on it. The estimation from the model is shown in red, which seems to match the data KDE reasonably well. In the right panel, we show the normalized covariance matrix for this model. Notably, our covariance matrices are dependent upon the model, which further goes into the computation of the likelihood for MCMC sampling.
 
 Next, we sampled the posterior distribution using the Monte Carlo Markov Chain (MCMC) method, more specifically, the Metropolis-Hastings algorithm \citep{1953JChPh..21.1087M}. We utilized the publicly available package \texttt{cobaya} \citep{2021JCAP...05..057T}\footnote{\url{https://cobaya.readthedocs.io/en/latest/}} to run the MCMC chains.  
We checked the convergence of the chains following the  Gelman-Rubin $R - 1$ statistic \citep{1992StaSc...7..457G}. The chain was assumed to be converged if the $R - 1$ value became less than a threshold $0.01$. 

 \begin{figure*}
    \centering
    \includegraphics[width=\textwidth]{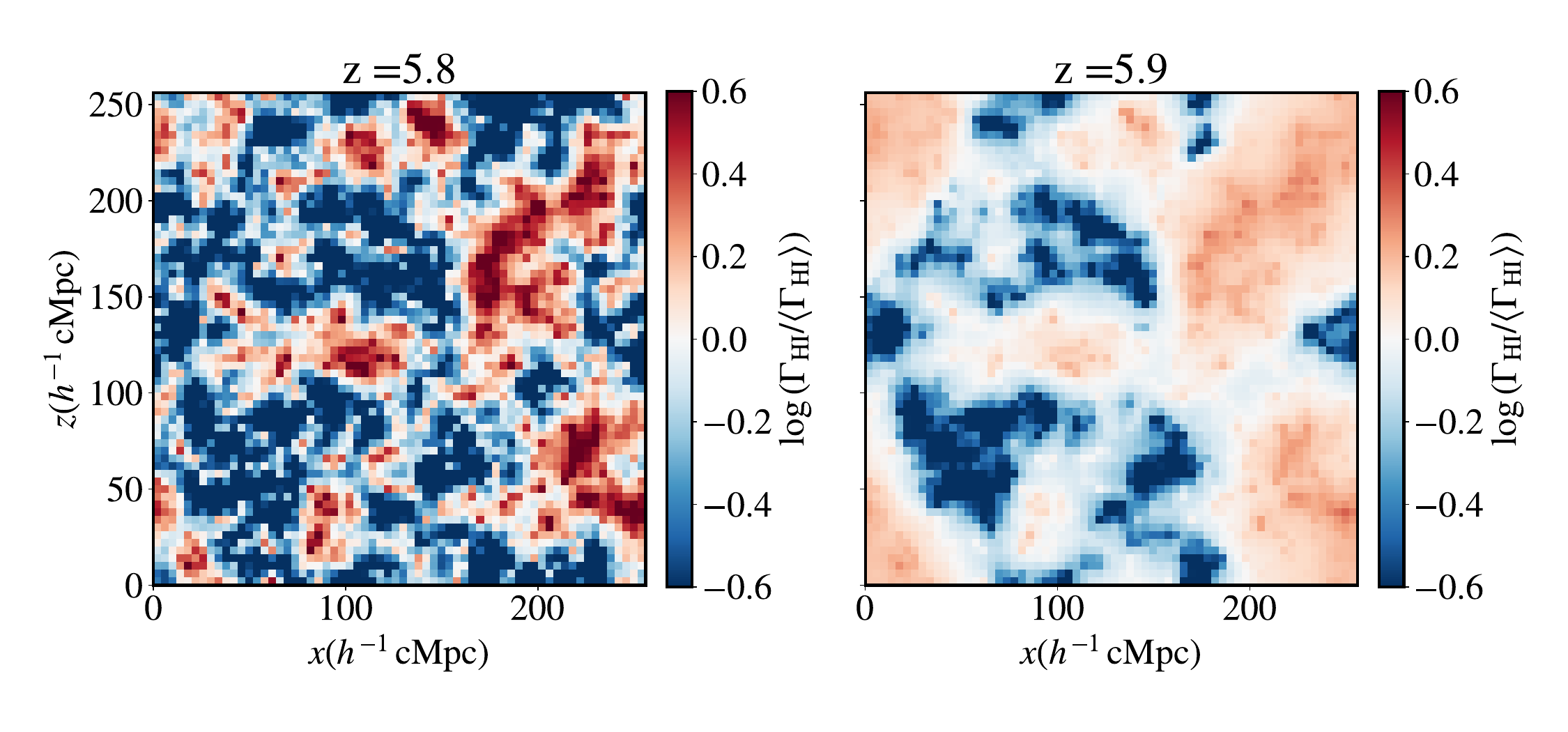}
    \caption{A comparison of UVB fluctuation fields allowed by the posteriors at redshift, $z=5.8$ (reionization is end) and $z=5.9$ (reionization is incomplete). The mean free path requirement for $z=5.8$ ($\lambda_0=4~h^{-1}\mathrm{cMpc}$) is much smaller than $z=5.9$ ($\lambda_0=40~h^{-1}\mathrm{cMpc}$), showing the effect of incomplete ionization on the UVB field. }
    \label{fig:fig12_Gamma_comp}
\end{figure*}

\begin{figure*}
\sidecaption
    \centering
    \includegraphics[width=0.7\textwidth]{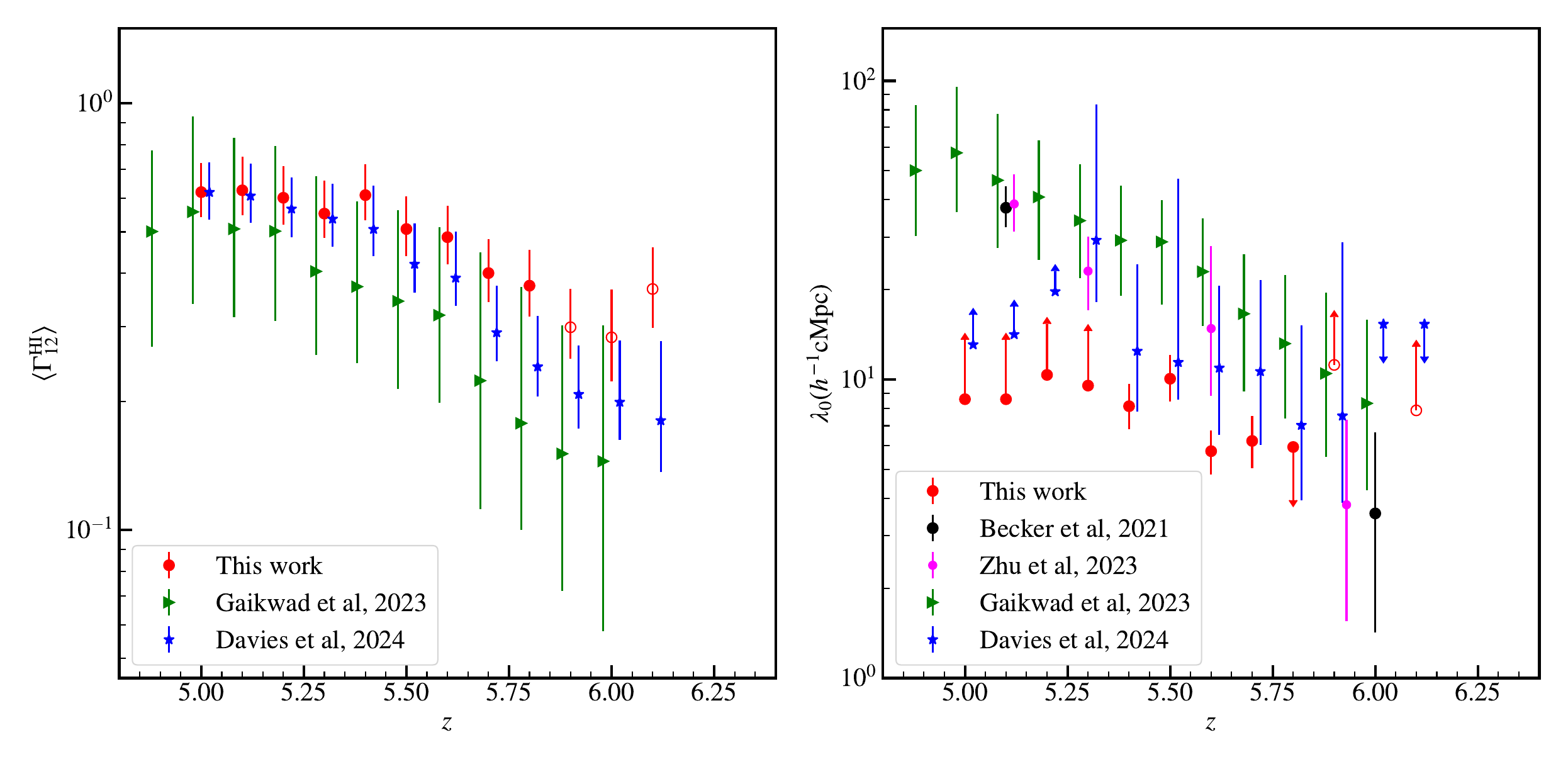}
    \caption{Marginalized 1D posterior constraints for and mean photoionization rate ($\langle\Gamma_{12}^{\mathrm{HI}}\rangle$) and  effective mean free path ($\lambda_0$) at different redshifts. The constraints correspond to 95\% confidence limits where the constraints are available from both sides. For limits, we quote the value where the posterior falls down to 1/$e^2$ from the maximum. The red circles denote the constraints derived from this work (hollow circles represent the redshift where reionization is not complete). The redshifts for green and blue points have been shifted by 0.02 dex for visual clarity.}
    \label{fig:fig13_1d_cons}
\end{figure*}

\section{Results \& interpretations}\label{sec:results}
In this section, we discuss the main results from our parameter space exploration studies using mean photoionization rate ($\langle \Gamma_{\mathrm{HI}}\rangle$) and effective mean free path ($\lambda_0$) as free parameters.
\subsection{Cumulative probability distribution of $\tau_{\mathrm{eff}}$}
We show the cumulative probability distributions (CDFs) of the effective optical depth ($\tau_{\mathrm{eff}}$) in Figure \ref{fig:tau_eff_ionfrac}, for the redshift range of our interest ($z=5.0-6.1$) in this study. The red lines show the bestfit CDF obtained by combining 500 random realizations of the sightlines while the number of sightlines is kept the same as provided by the observational data. The cyan curves in the background represent the individual realizations of the bestfit model. For the data (shown in black dashed lines), we show both optimistic and pessimistic limits, although we used only the optimistic case for the parameter space exploration. We checked that the bestfit evolution model can nicely match the observed scatter in the optical depth distribution as well as the mean transmission flux in the majority of the cases with mild deviation at $z=5.6$ and $5.9$. The deviation is mainly due to the dominance of $\tau_{\mathrm{eff}}$ values at the distribution of the edges in the data for those redshifts. Furthermore, the width of the distribution increases as we go towards higher redshifts, suggesting increasing fluctuations in the opacity fields. 
\subsection{Joint parameter posteriors}\label{subsec:joint_post}
Next, we quantify the posterior distribution of the parameters extracted from three different temperature models allowed by the current constraints on the IGM temperatures, as shown earlier in Figure \ref{fig:fig5_fid_model}.  We perform this analysis for all the redshifts of interests and show the combined posteriors in Figure \ref{fig:fig10_Gamma_lambda_mcmc_posterior} while in Figure \ref{fig:fig11_2d_post_demons}, we show the 2D posterior contours (at 68\% and 95\% confidence level) for the individual temperature models along with the combined contour of all the three samples again at the fiducial redshift, $z=5.4$.  It is apparent that the combined contour nicely represents the whole posterior space allowed by the temperature constraints. Now, moving the discussion to Figure \ref{fig:fig10_Gamma_lambda_mcmc_posterior} for different redshifts, it can be seen that the data are consistent even with almost uniform UVB models towards low redshifts ($z\le5.3$), demanding only a small amount of UVB fluctuations. This interpretation is driven by the fact that, in our definition, post-reionization UVB fluctuations are solely quantified by the mean free path parameter $\lambda_0$. The large parameter values correspond to a relatively uniform UVB field and vice versa. As at those redshifts, all the posteriors hit the upper end of the prior values on $\lambda_0$, we can conclude that almost uniform UVB models are also allowed by the data. An interesting feature can be noticed at  $z=5.3$ where a bimodality in mean free path distribution arises disfavouring some of the moderate UVB fluctuation models ($\lambda_0\sim20-30 h^{-1}\mathrm{cMpc}$). This happens because those models are not able to provide enough fluctuations to match the data distribution. For a large mean free path, the opacity fluctuations are driven by the underlying density field and temperature fluctuations. As the mean free path decreases, it initially neutralizes the opacity fluctuations as demonstrated in Figure \ref{fig:fig7_tau_snapshots}, which corresponds to the disfavored region in the posterior. This happens because high-density regions are ionized earlier and have relatively lower temperatures at later times due to the availability of more time for cooling \citep{2015ApJ...813L..38D}. This enhances the opacity fluctuations correlating with density fluctuations. However, the correlation between density and UVB field is opposite, providing high photoionization rate at high density regions and vice versa, which further drives the fluctuations in the other direction \citep[see also; ][]{2018ApJ...860..155D}. Then, with a reasonably short mean free path, the opacity fluctuations are dominated by the UVB field itself. Hence, the bimodal feature portrays the interplay among these competing fluctuations, generating the opacity field.  

As we move towards higher redshifts (up to $z=5.8$, where reionization ends in our model), the required UVB fluctuations also increase, allowing for shorter mean free paths. Before reionization ends ($z \ge 5.9$), the ionization fluctuations (driven by leftover neutral islands) are strong enough to produce the required large-scale UVB fluctuations matching the observed opacity fluctuations, hence requiring little or almost no fluctuations driven by post-reionization $\lambda_0$. This further keeps the mean free path parameter unconstrained. This explanation can be clearer from Figure \ref{fig:fig12_Gamma_comp}, where we show two snapshots of UVB fluctuations. Among these two snapshots, one is at redshift, $z=5.8$ (after reionization completion) with a very short mean free path ($\lambda_0=4~ h^{-1}\mathrm{cMpc}$) and the other one is at redshift, $z=5.9$ (before reionization completion) with a relatively high mean free path parameter ($\lambda_0=40~ h^{-1}\mathrm{cMpc}$), as suggested by the posteriors. It is apparent that the presence of neutral regions can produce strong fluctuations in the UVB field even if $\lambda_0$ is larger \citep[see also, ][]{2021MNRAS.501.5782C}. This supports our above argument for an unconstrained mean free path at higher redshifts where reionization is still not ended. Hence, despite the fact that the $\lambda_0$ posteriors look very similar to the cases at $z\leq5.3$, the UVB fluctuations are not uniform here (at $z=5.9-6.1$), rather driven by the leftover neutral islands. This analysis provides us with the confidence that our simplistic model can be utilized to pursue parameter space statistics study and interpret the relevant physics from the observational data.
\subsection{constraints on $\langle\Gamma_{12}^{\mathrm{HI}}\rangle$ and $\lambda_0$}
In Figure \ref{fig:fig13_1d_cons}, we show the marginalized constraints on the mean photoionization rate ($\langle\Gamma_{12}^{\mathrm{HI}}\rangle$, at the left panel) and effective mean free path ($\lambda_0$, at the right panel) at 95\% confidence level along with the points from recent literature. Our estimates are denoted with red circles (the solid ones indicate the redshifts where reionization is completed, and hollow ones indicate the redshifts with incomplete reionization). We found that our photoionization rate estimates are consistent with the findings of \citet[][\textit{shown in blue stars with displaced redshift for better visualization}]{2024ApJ...965..134D}. This is not very surprising, as the UVB fluctuation generation model is the same for both cases. However, our points are slightly higher as we have a realistic reionization history allowing sufficient time for cooling in the early ionized high-density regions. On the other hand, \citet{2024ApJ...965..134D} assumed a sudden reionization for all the cells in the simulation box, which results in larger temperature and hence, a slightly shorter photoionization rate to match the mean transmission flux. Additionally, our estimates are also in agreement with the corresponding constraints of \citet[][\textit{shown in green triangle again shifted in redshift}]{2023MNRAS.525.4093G}. 

In the case of the mean free path, we are able to derive the constraints from posteriors at redshift range $z=5.4-5.7$. For the redshifts providing constraints from only one side, we quote the value where the posterior value falls down to $1/e^2$ from the maximum \citep{2024ApJ...965..134D}. We do not get any constraints at $z=6.0$. Our constraints prefer a shorter mean free path distribution than \citet{2024ApJ...965..134D} values, although these are consistent within the error bars, particularly given that the lower end of the \citet{2024ApJ...965..134D} uncertainties corresponds to an approximate correction to account for their lack of model self-consistency. The main reason for this difference is again traced back to the assumed sudden reionization and the lack of post-reionization large-scale temperature fluctuations in their model.  It is expected that the large-scale temperature field is anticorrelated with the density field due to the variations in reionization timing at different regions. As a result of that, a stronger UVB fluctuation is required to overcome the correlated fluctuations of thermal and density field, which in turn demands a lower $\lambda_0$ value. Intriguingly, this more physical treatment with a gradual and patchy thermal evolution puts our model constraints within a similar ballpark of recent high-redshift mean free path measurements \citep{2021MNRAS.508.1853B, 2023ApJ...955..115Z}. At this point, it is important to highlight that the definition of our mean free path is similar to \citet{2024ApJ...965..134D}, which is solely a subgrid parameter quantifying clumping of the unresolved gas. Hence, our inferred mean free path only characterizes the excess UVB fluctuations over the uniform case without including density field fluctuations. As a consequence, we do not get any upper bound on $\lambda_0$ for $z\le5.3$ as an almost uniform UVB field is consistent with fluctuations required by the data. This is unlike the approach adopted by \citet{2023MNRAS.525.4093G}, where the mean free path is computed along with their density field after including fluctuating UVB contribution. That further enables them to put constraints on the mean free path even without requiring explicit UVB fluctuations. At redshift $z\geq5.9$, reionization is incomplete in our model, which further allows us to provide only lower limits on the mean free path, as demonstrated earlier in the subsection \ref{subsec:joint_post}.  The constraints used in the plots are tabulated in Table \ref{tab:param_cons}.

In continuation with the above discussion, it is worthwhile to mention that most of the semi-numerical models give a sharp transition to the reionization end, while detailed numerical simulations \citep{2022MNRAS.511.4005K,2025MNRAS.539L..18A} prefer relatively smooth transitions along with slight inflection in neutral fraction evolution. A gradual transition will translate to a smoother evolution in mean free path, as hinted from our findings. A smooth transition towards reionization end can be achieved within a semi-numerical setup with more realistic modelling of photon sinks (e.g. \citealt{2022MNRAS.514.1302D,2025PASA...42...49Q}). This can provide a direction for potential improvement for our model in the future.
\begin{table}
\centering
\renewcommand{\arraystretch}{1.2}
\setlength{\tabcolsep}{7pt}
\begin{threeparttable}
\begin{tabular}{ccccc}
\hline
Redshifts ($z$) & \multicolumn{2}{c}{$\langle \Gamma_{\mathrm{HI}}^{12}\rangle$} & \multicolumn{2}{c}{$\lambda_0 (h^{-1}\mathrm{cMpc})$} \\
\hline  
& Mean & 95\% Limits & Mean & 95\% Limits \\
\hline
5.0 & 0.62 & [0.54, 0.73] & -& >8.59\\
5.1 & 0.63 & [0.55, 0.75] & -& >9.12\\
5.2 & 0.60 & [0.52, 0.71] & -& >10.36\\
5.3 & 0.55 & [0.48, 0.66] & -& >9.54\\
5.4 & 0.61 & [0.53, 0.72] & 8.14 & [6.81,9.69] \\
5.5 & 0.51 & [0.43, 0.66] & 10.05 & [8.44, 12.05]\\
5.6 & 0.49 & [0.42, 0.58] & 5.75 & [4.80, 6.76]\\
5.7 & 0.40 & [0.34, 0.48] & 6.22 & [5.04,7.53]\\
5.8 & 0.37 & [0.31, 0.45] & -& <5.94\\
5.9 & 0.30 & [0.25, 0.37] & -& >11.19\\
6.0 & 0.28 & [0.22, 0.37] & -& -\\
6.1 & 0.37 & [0.30, 0.46] & -& >7.88\\
\hline

\end{tabular}
\end{threeparttable}
\caption{Parameter constraints obtained from the MCMC-based analysis after combining different temperature models. For each redshift, we show the mean value of combined samples along with 95\% confidence limits on different parameters where constraints are achieved from both sides. We quote the value where the posterior value falls down to $1/e^2$ from the maximum \citep{2024ApJ...965..134D} for one sided limit.}
\label{tab:param_cons}
\end{table}
\section{Summary \& conclusions}\label{sec:summary}
The large-scale fluctuations of Ly-$\alpha$ forest spectra observed in high-redshift quasars contain crucial information about the physical properties of the IGM. The modelling of these fluctuations becomes challenging due to the effects of various physical quantities at different scales. Ideally, one would require an extremely high dynamic range simulation in order to accurately characterize the fluctuations. However, this becomes computationally intractable given a wide range of uncertain astrophysical parameters. To this end, fast semi-numerical approaches can show us an alternate way to provide an optimization between speed and accuracy.   

In this work, we developed an efficient semi-numerical model to estimate the Ly-$\alpha$ opacity fluctuations during the late and post-EoR phase. We further compared our model with the available observational data, aiming to constrain the model parameters. Below, we summarize the key points of this work.
\begin{itemize}
    \item We calibrated our model with respect to high resolution full hydrodynamic Nyx simulation. This essentially modified the fluctuating GP relation accordingly to be applicable on coarse resolution grid cell (in our case $4~h^{-1}\mathrm{cMpc}$). Our model self-consistently included the relevant ingredients (i.e., density, temperature, and UVB fluctuations) for generating the opacity fluctuations. We checked that the opacity distribution using the calibrated model can provide a reasonable match with the full hydro skewers. The small differences may arise due to the lack of velocity field correlation in our model. However, this effect is unlikely to be strong and can be explored in a separate future project.
    \item Next, we demonstrated the capabilities of our model assuming fiducial ionization and thermal histories that are consistent with the available constraints.  We found that, along with the underlying density field, large-scale temperature fluctuations can also play a role in shaping the opacity fluctuations. Further, the photoionization rate/UVB fluctuations act oppositely to the above two. Hence, one needs a very short effective mean free path ($\lambda_0$) to produce sufficient opacity fluctuations after overcoming the density and temperature contribution. We also found that the mean photoionization rate ($\langle\Gamma_{12}^{\mathrm{HI}}\rangle$) mainly controls the mean transmission flux by shifting the opacity distribution.
    \item We then compared our model predictions with the opacity fluctuations measured by the extended XQR-30 dataset. The efficiency of our model also allowed us to perform a parameter space exploration study by varying $\lambda_0$ and $\langle\Gamma_{12}^{\mathrm{HI}}\rangle$. We found that the bestfit model, along with the scatter due to realization, can reasonably match the data distribution. This further enabled us to derive constraints on these parameters. The constraints remained consistent within the error bars with the previous measurements.
\end{itemize}
Given the prospective wealth of high-quality spectroscopic data at high redshifts in the near future, it is crucial to have efficient and reasonably accurate models for gleaning relevant physical information without major computational bottlenecks. To this end, our model can be useful to study Ly-$\alpha$ opacity fluctuations in a large cosmological simulation volume within a tractable computational cost. Hence, this model can complement the full-scale hydrodynamic simulations when there is a requirement for uncertain parameter space exploration.  With the availability of upcoming large field-of-view telescope surveys, this will be crucial from a theoretical perspective. Furthermore, this can be useful for the synergetic multi-wavelength study of EoR along with other probes like the 21 cm signal from neutral hydrogen distribution. We plan to exploit these capabilities of our model in future projects.


\section*{Data Availability}

The data presented in this article will be shared on reasonable request to the corresponding author (BM).

-------------------------------------------------------------------
\bibliographystyle{aa}
\bibliography{aa55515-25}
\appendix
\section{Justification for Gaussian likelihood}
Here, we show a plot justifying the choice of our Gaussian likelihood for the parameter exploration studies. The plot shows the scatter between two random bins due to different realizations of the sightlines, along with a few Gaussian contours. The nature of the scatter is reasonably close to being Gaussian, which gives us the confidence to use a standard multivariate Gaussian likelihood in this study.
\begin{figure}
    \includegraphics[width=0.5\textwidth]{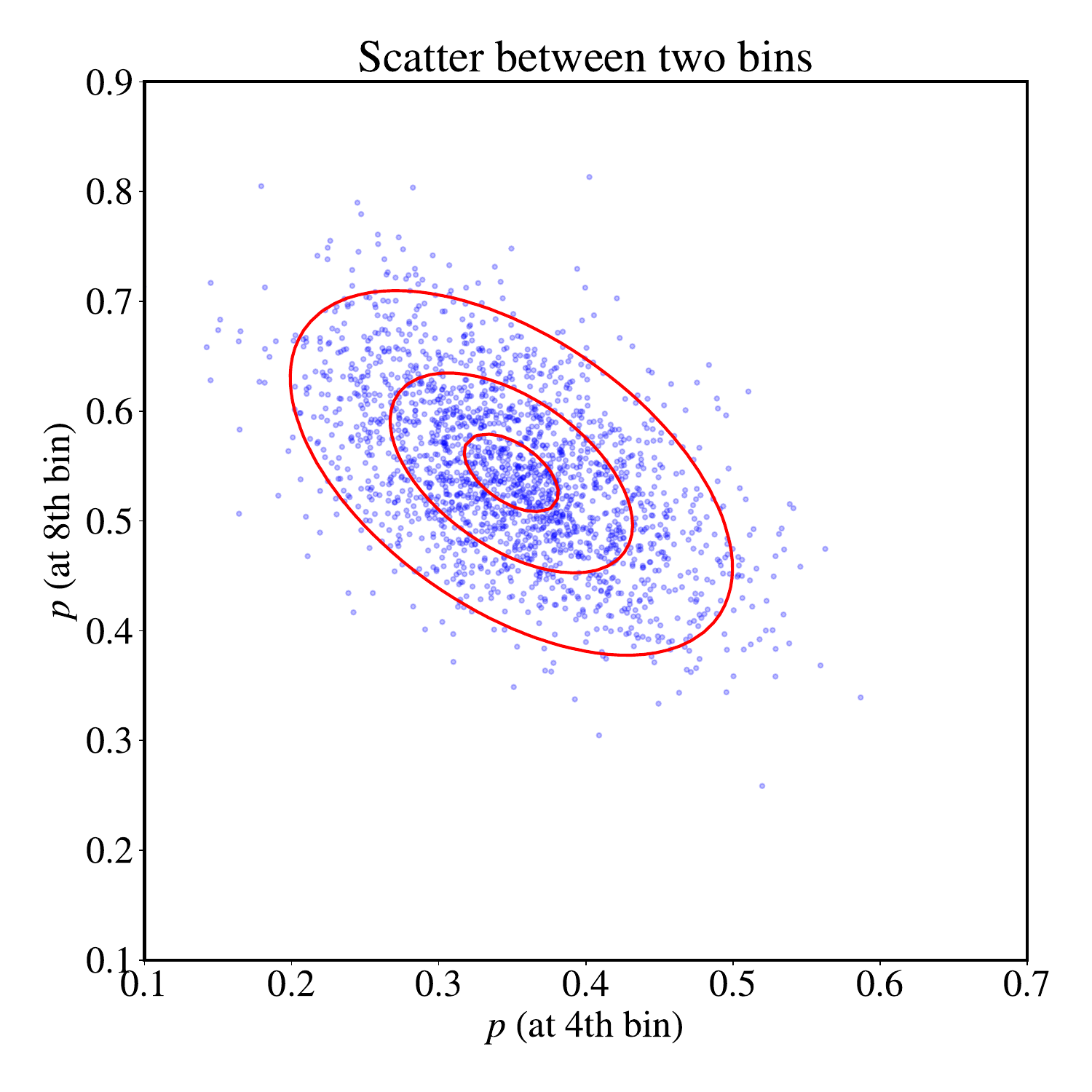}
    \caption{The scatter between two bins (specifically 4th and 8th). }
\end{figure}
\end{document}